\documentclass[final,5p,times,twocolumn]{elsarticle}
\usepackage{amsmath,amsfonts}
\usepackage{algorithmic}
\usepackage{array}
\usepackage[caption=false,font=normalsize,labelfont=sf,textfont=sf]{subfig}
\usepackage{textcomp}
\usepackage{stfloats}
\usepackage{url}
\usepackage{verbatim}
\usepackage{graphicx}
\usepackage{cleveref}

\usepackage{siunitx}
\usepackage{balance}

\newcommand{\MSun}{\mbox{${M}_\odot$}}

\def\aplt{\ {\raise-.5ex\hbox{$\buildrel<\over\sim$}}\ }

\newcolumntype{R}[1]{>{\raggedleft\arraybackslash}p{#1}}

\def\aap{\ {A\&A}\ }

\def\aj{\ {AJ}\ }

\def\apj{\ {ApJ}\ }
\def\apjl{\ {ApJL}\ }

\def\apss{\ {Ap\&SS}\ }

\def\mnras{\ {MNRAS}\ }

\def\pasj{\ {Publ. Astr. Soc. Japan}\ }

\begin{document}
\begin{frontmatter}

  \title{Validating direct solvers for Newton's gravitational N-body problem, and the systematic comparison between IEEE floating point and Posits.}
  
\author{Simon Portegies Zwart\footnote{E-mail: spz@strw.leidenuniv.nl}
\affiliation{Leiden Observatory, University of Leiden, Einsteinweg 55, 2333 CC, Leiden, The Netherlands}
}
\date{Accepted XXX. Received YYY; in original form ZZZ}

\begin{abstract}
We present a systematic comparison between arbitrary precise
arithmetic and integration, IEEE-754 compliant floating point
arithmetic (fp16, bfp16, fp32, double precision fp64, and quadruple
precision fp128), and two implementations of Posits (type III unum)
for solving Newton's chaotic N-body problem. Each implementation is
benchmarked with arbitrary precise calculations to objectively
evaluate their performance in precision as well as speed. We rely on
hardware and compiler implementations for fp64, and software
implementations for arbitrary-precision arithmetic and Posits. Half
precision arithmetic (fp16, bfp16, and Posits$<16,1>$) are
insufficiently precise for solving Newton's equations of
motion. Single precision (fp32, and Posits$<32,2>$) could be used for
statistical ensemble calculations, but lead to relatively large errors
in any individual strong encounter. All 64-bit implementations fp64 as
well as Posits (Posits$<64,3>$) experience difficulty in our
tests. One of the implementations of Posits (Universal) gives
precision comparable to fp64 but is slow (by at least an orders of
magnitude compared to fp64 after correcting for the more efficient
hardware support for the latter). The other (CPPPosits) has a speed
comparable to fp64 but has systematically larger errors (by about an
order of magnitude compared to fp64 with excesses exceeding two orders
of magnitude). As a consequence, this implementation leads to a
systematic drift in the result space and has difficulty resolving
close encounters. Posits and fp64 have difficulty when integrating a
dynamical system in a moving reference frame; testing Galileo
invariancy. In their current implementation, Posits do not seem to be
the ideal alternative for fp64 when integrating chaotic or stiff
ordinary differential equations, such as Newton's equations of motion.
\end{abstract}


\end{frontmatter}

\section{Introduction}\label{sec:intro}

Floating point arithmetic has been around well before the first
digital computers. Konrad Zuse's Z3 computer, completed in 1941, was
the first programmable digital computer with 22-bit binary
floating-point arithmetic unit and built-in exception handling. Since
then, floats have improved substantially and form the standard
throughout scientific computing. Numerical floating-point standard,
called IEEE-754 representation, defines layouts for the sign,
exponent, and significance (mantissa) for half (16-bit), single
(32-bit), double (64-bit, hereafter fp64), and quadruple (128-bit)
precision \cite{IEEE754-2019}.

Recently, John Gustafson designed a new precise arithmetic
representation for digital computers called unum (for universal
number) \cite{gustafson2015end}. After two generations of development,
unum was replaced by Posits (type III unum)
\cite{Gustafson2017Posit,PositStandard2022}\footnote{The Posit
standard can be found at
\url{https://posithub.org/docs/posit_standard-2.pdf}}. Posits offer
domain scientists a novel way to represent floating-point numbers for
general-purpose computing. The advantages of Posits have been
acclaimed, in particular for speed, error handling, and dynamic
range. The advantages of Posits over float are anticipated to
\begin{itemize}
\item[$\bullet$] achieve higher precision than floats for the same
number of bits. This would be particularly noticeable for numbers
near unity \cite{ijcer2022posit}.
\item[$\bullet$] dynamically adapts precision, and therewith provides
greater dynamic range for the same number of bits \cite{vuthaj2023from}.
\item[$\bullet$] requires fewer and simpler operations for achieving similar precision,
making a potential hardware implementation simpler \cite{Ciocirlan2021}.
\item[$\bullet$] support the "quire" accumulator for exact dot
  products \cite{2024arXiv240114117N}.  Note that fp64 can have an
  exact accumulator via a Kulisch accumulator
  \cite{kulisch2008computer}, but posits standardized this into the
  design \cite{9817027M}.
\item[$\bullet$] and in contrast to fp64, Posits have a unique
  representation for zero.  Rather than a speclial bit patterns for
  infinities and NaNs, Posits throw one special value, NaR ("Not a
  Real") for mathematically invalid operations, such as division by
  zero. Instead of underlow or overflow, as fp64, they round to the
  largest finite posit or to the smallest non-zero posit
  \cite{Gustafson2017Posit}.
\end{itemize}
The apparent advantages of Posits sound wonderful for any numerical
scientist, and therefore, it is time to put them to the test.  Earlier
work compares the performance of Posits for IEEE-754 floating point
representation using 32 bits \cite{FernandezHart2024Izhikevich}, and
other length \cite{10.1145/3316279.3316285}. They found no performance
difference between 32-bit Posits and 32-bit floating-point, but 16-bit
Posits outperform 16-bit floating point for regular tonic
spiking\footnote{A numerical representation of neural activity. In
this specific case, they discuss where artificial neurons fire at a
consistent rate, without bursts or irregular intervals.} by a factor
of 18 \cite{FernandezHart2024Izhikevich}.  \cite{COVENEY2024102449}
(see their Fig.2) briefly experimented with Posits in reproducing
chaotic generalized Bernoulli map, but their results for Posits were
not optimistic. \cite{Guthmuller2025VRP} arrived at a similar
conclusion, claiming that posits were insufficiently precise in
comparison with floats of the same length.

As an application, we use the integration of Newton's equations of
motion for 3 and 100 bodies of comparable mass. This problem is
excellently suited for a rigorous numerical test of Posits in
comparison with floating points because the problem is sensitive to
small errors \cite{1891BuAsI...8...12P} and covers a wide dynamic
range capped with under- and overflow \cite{2020gfbd.book.....M}.

\section{The experimental setup}

We design a series of N-body solvers to integrate Newton's equations
of motion of N point masses in time. These calculations start with an
initial realization of the system in 6N-dimensional phase space
(assuming equal and constant masses), which we call the initial
conditions. Those conditions were generated in IEEE-754 double
precision and stored in a file. All runs start with the same initial
conditions file.  Lower-precision arithmetic represents numbers with
fewer digits, so values are rounded to the nearest representable
number whenever they cannot be stored exactly: For example when inputs
are converted to the lower precision or after each arithmetic
operation.  The 128-bit initial conditions extend the initial
conditions with zeros up to the least significant digit.  The
conversion of the fp64 initial conditions to another format introduces
at most a single rounding error per coordinate, which is of the same
order as the round-off made in each of the $\sim 10^9$ or more
subsequent arithmetic operations. Alternatively, one could have
generated initial conditions that are exactly representable in the
lowest-precision format, so that all arithmetic starts from the same
state, or initial conditions more precise than any of the adopted
formats, so that the conversion error affects all formats alike.  We
do not expect either choice to affect our conclusions, because the
differences we find between the formats are driven by the accumulated
round-off during the integration, which is subsequently magnified
exponentially by the chaotic nature of the problem (see
\cref{Sect:Chaos}), rather than by the initial conversion.

We employ two very different N-body codes, one for validation (which
we describe in \cref{Sect:Brutus}), and one for conducting the
experiments (described in \cref{Sect:Hermite}).

\subsection{The consequences of chaos on reproducibility}\label{Sect:Chaos}

The gravitational N-body problem is chaotic for $N>2$, and the results
are susceptible to small errors. This poses considerable complications
in direct code comparisons because small deviations from the correct
(converged) solution (such as round off errors) cause the system to
behave differently. This behavior makes it difficult to design a
validation and verification test.  Result can be validated by
comparing them to a more precise and accurate solution, and quantify
the moment the two solutions deviate.  A more qualitative validation
can be achieved by monitoring the conserved quantities, time
reversibility, and by characterizing the system's chaotic behavior.

The strict conservation of energy, linear momentum, and angular
momentum are fundamental requirements for realism and reliability in a
classical gravitational simulation. Any deviation from these conserved
quantities render the result fundamentally unreachable for the adopted
initial conditions (masses, positions, and velocities). Still, in
numerical simulations, we allow for infinitesimal deviations from
these conserved quantities, even though we do appreciate their strict
conservation.

So long as the system responds linearly to infinitesimal deviations in
phase space, assessing the quality of the simulation is relatively
straightforward. The classical few-body problem, however, is subject
to bounded chaos, resulting in a non-linear (typically exponential)
response to perturbations. These perturbations may come from
integration errors, and round off.

Any deviation introduced affects the conserved quantities and
renders the solution nonphysical. Even when these deviations remain
below some pre-determined threshold, the actual phase space solution
exponentially deviates from the true solution. As a consequence, we
have to deal with two independent errors: errors in the conserved
quantities and deviations from the shadowed phase space
characteristics.

The first variety of errors we can quantify by measuring the deviation
of the conserved quantities. If the chaos-driven exponential growth of
such errors exceeds the error-baseline, the final realization of the
system is governed by the initial error. The final phase-space of such
a simulation is still expected to cover the true outcome space of the
initial conditions, modulo an infinitesimal perturbation. We refer to
these as {\em nagH hoch} (or case-A errors, for Acceptable solutions)
\cite{2018CNSNS..61..160P}.  However, if the baseline error per
operation exceeds the chaotic growth of earlier errors, the final
phase space coverage becomes apprehensive. The results of such
simulations do not represent the final phase-space coverage of the
actual dynamical system. We call these case-B errors (for Bad
solutions).

The second variety is harder to quantify. One of the complications
arises when the difference between the conserved quantities of two
calculations remains below the acceptable threshold, but their
phase-space distance (see \cref{Eq:phase_space_distance}) grows. In a
chaotic system, such as Newton's gravitational N-body problem, the
phase-space distance between two solutions grows exponentially. We
measure this growth through the magnification factor ($\alpha$), and a
time scale. When the magnification factor $\alpha \equiv e$, we call
this time scale the Lyapunov time scale, or $t_{\lambda}$. We quantify
the system's sensitivity to perturbations by measuring $t_{\lambda}$.
Such analysis has to stop well before the phase-space distance between
two solutions becomes on the order of the size of the system. By then,
both systems have deviated beyond recognition, and their states should
not be compared.

Researchers often silently hope (or maybe even expect) that a
simulation complies to {\em nagH hoch}. Parameter space is
subsequently covered by performing ensemble averages, they hope
(and expect) that the chaotic behavior of the system can be described
with Poissonian statistics. In that case, an ensemble average can be
interpreted as a probability density distribution in outcome
space. This statistic, however, is rarely validated, and potentially
inappropriate. As a consequence, one should not rely on the results of
a single calculation of a chaotic process, but statistically analyzing
an ensemble average may not be appropriate either.

Case B errors compromise the credibility of any simulation; The
probability distribution of the simulation's outcome space only
partially represents the true outcome space, with some regions of the
actual outcome space remaining unrepresented. As a consequence, it is
not possible to assess the simulation's reliability.

We quantify the first variety of errors by performing a phase-space
comparison to a converged solution obtained through a validation
calculation (see \cref{Sect:Brutus}). This comparison allows us to
quantify the consequences of integration and round-off errors. We
quantify the second variety of errors by measuring $t_{\lambda}$. Both
quantifications require a converged solution of the time evolution of
the Newtonian system.

\subsection{Validation calculations}\label{Sect:Brutus}

We compare the results of floats and Posits with converged solutions,
calculated using arbitrary-precision floating-point arithmetic. A
solution to the N-body problem that has converged to the $n$-th
decimal is a solution for which the $n$-th decimal place is unaffected
by numerical precision or time-step errors. Acquiring such a solution,
even for three decimal digits of accuracy ($n=3$), is non-trivial and
requires a special numerical setup.

We adopt the Brutus arbitrary-precision N-body code
\cite{2015ComAC...2....2B}. Brutus is designed for simulating the
gravitational N-body problem with exceptional high numerical precision
and accuracy. This is achieved by combining arbitrary-precision
arithmetic through MPFR
\cite{Fousse:2007:MMB:1236463.1236468}\footnote{An interesting
alternatively to MPFR, could be using xvpfloats
\cite{Guthmuller2024Xvpfloat} which are claimed to give good
performance up to 512-bit arithmetic.} with the Bragg-Bulirsch–Stoer
integrator \cite{springerlink:10.1007/BF01386092} using adaptive
time-stepping to acquire the necessary accuracy. The precision of
Brutus is then tunable by specifying the number of bits for the
arithmetic operations (64 bits for double precision, but here only
limited by memory and patience), and a tolerance parameter
($\epsilon$) to control the accuracy of the Bulirsch–Stoer integrator.
By systematically varying the arithmetic precision (number of bits)
and the integrator tolerance ($\epsilon$) we iteratively perform the
calculation with the same realization until the solution converges to
a specified (pre-determined) number of significant digits.

\subsection{The numerical Newtonian N-body toolbox}\label{Sect:Hermite}

For our experiments, we adapt the direct N-body code {\tt GravityLab}
\cite{1995ApJ...443L..93H}, and generalized it for the use of various
bit-length IEEE-754 floating point arithmetic and Posits. The code
integrates Newton's equations of motion using the 4th order
predictor-corrector Hermite scheme \cite{1992PASJ...44..141M}. It
employs dynamic time steps which are calculated per step using the
smallest free-fall time scale at the current realization of the
particles, and subsequently multiplied by a tuning parameter $\eta$.
We call this code {\tt NBotox}, and it is publicly available at
\url{https://gitlab.strw.leidenuniv.nl/spz/NBotox}.  Calculations are
performed in dimensionless N-body units \cite{1971Ap&SS..13..284H},
except for the case of D9 in \cref{Sect:Results.D9}, for which we
adopt astrophysical units.

\subsection{The Posits implementation}\label{Sect:Posits}

We compare two independently developed implementations of Posits with
floats of various bit-lengths. Both implementations are written as C++
template libraries. Here we adopt standard Posits with 1, 2, and 3
bits for the exponent depending on the wordlength, or posit$\langle
16,1 \rangle$, posit$\langle 32,2 \rangle$, and posit$\langle 64,3
\rangle$.

The one implementation is taken from
\url{https://github.com/federicoops/cppposit?tab=readme-ov-file}
(version v1.0.0), which is a fork from
\url{https://github.com/unipi-dii-compressedarith/cppposit}, which again is based on an earlier implementation by
\cite{Gustafson2017Posit}.
We refer to this implementation as {\rm CPPPosits}.

The other is part of the {\tt Universal} library developed in
\cite{omtzigt2023Universal}, which is based on earlier work by
\cite{10.1007/978-3-031-09779-9_7}. We adopted the implementation
presented in
\url{https://github.com/stillwater-sc/Universal?tab=readme-ov-file}.
We refer to this implementation as {\em Universal}.

\subsection{The difference between Posits (unum type-III) and universal implementations}\label{Sect:Posits_vs_Uiveral}

Both Posit libraries are implemented in C++ but they differ in scope,
design philosophy, and how they treat the broader Unum ecosystem.  The
CPPPosit library specializes, making it well-suited for usage in
relatively simple codes, but as a result it is less flexible.  The
Universal library, on the other hand, provides an ecosystem of
alternative arithmetic implementations.  Both libraries are
implemented as C++ (C++11/14) templates (and rather header-heavy), and
claim to implement Gustafson's Unum Type-III (Posit); still we find
profound differences in performance, precision, and accuracy.

The main implementation difference is probably the use of a tabulated
fixed-point back-end (big-integer style, 64/128‑bit) in CPPPosits,
whereas in the Universal library this is part of the core design.
In Universal, the bits of a number are stored in a generic container
(a so-called block-binary) that consists of an array of machine
integers (limbs; for example 8, 32, or 64-bit unsigned integers).
Arithmetic on such a number is performed limb by limb using integer
operations, including the explicit propagation of carries between
limbs, in the same way as multiple-precision integer libraries do.
This makes it possible to support posits of arbitrary length and
exponent size with the same code, but every floating-point operation
is then reduced to a sequence of integer operations on these limbs.

\section{The Experiment}

We set up three 3-body experiments, and one 100-body experiment. The
first two experiments are planar and selected for their high
sensitivity to one of the specific errors common in integrating N-body
systems. The first experiment is sensitive to variations in orbital
phase; a quantity that is not preserved in the laws of physics. The
second experiment is sensitive to errors in energy and angular
momentum. The third 3-body experiment, motivated by a recently
discovered binary orbiting a super-massive black hole, is
3-dimensional and sensitive to secular time-scale angular momentum
exchange between the inner and the outer orbit.  We selected this
experiment for the extended dynamic range needed to effectively solve
the numerics.

For the first experiment, we adopt the gravitational 3-body braid
\cite{0951-7715-11-2-011} (The so-called figure-8 orbit.). This is a
periodic orbit where three stars orbit each other along the same
trajectory with different constant relative orbital phases. The orbit
is linearly stable \cite{2026A&A...707A.215P}. It is relatively
insensitive to energy errors but sensitive to small errors in angular
momentum and orbital phase. Note that orbital phase is not a conserved
quantity in a Newtonian self-gravitating system, and it may require a
symplectic scheme to resolve the subtleties of the dynamics
\cite{2003A&A...400.1129B}, which our adopted numerical integrator is
not.

For the second experiment, we adopt the Pythagorean 3-body
problem, introduced by Burrau in 1913 (But already investigated by
Ernest Meissell in 1893, and solved by Szebehely in 1967). We adopt
the Pythagorean 3-body problem here because it remains a classic
example in celestial mechanics \cite{1994CeMDA..58....1A}. Despite its
short lifetime, the Pythagorean problem is hard to solve, providing an
excellent experiment for numerical rigor.

The third experiment tests the secular stability of the integrator in
extreme dynamic range.  In \cref{tab:InitialConditions}, we present
the initial conditions for the figure-8, the Pythagorean problems,
and the integration of the binary D9 in orbit around the supermassive
black hole.

\begin{table*}[h]
\centering
\caption{Initial conditions for the figure-8 problem
\cite{0951-7715-11-2-011}, and the Pythagorean problem. The
parameters are presented in N-body units
\cite{1971Ap&SS..13..284H}. Initial conditions are presented in 8
decimal places: values of $0$ in input are equivalent to
$0.0\rlap{0}/$. }
\begin{tabular}{
S[table-format=1.0]
S[table-format=1.9]
S[table-format=1.9]
S[table-format=1.0]
S[table-format=1.9]
S[table-format=1.9]
S[table-format=1.0]
}
\hline \hline
&\multicolumn{3}{l}{position} & \multicolumn{3}{l}{velocity} \\
\hline
\multicolumn{4}{l}{figure-8} \\
$m$ & $x$ & $y$ & $z$ & $v_x$ & $v_y$ & $v_z$ \\
\hline
1 & 0.97000436 & -0.24308753 & 0 & 0.466203685 & 0.43236573 & 0 \\
1 & -0.97000436& 0.24308753 & 0 & 0.466203685 & 0.43236573 & 0 \\
1 & 0 & 0 & 0 & -0.932407370 & -0.86473146 & 0 \\
\hline
\multicolumn{4}{l}{Pythagorean} \\
$m$ & $x$ & $y$ & $z$ & $v_x$ & $v_y$ & $v_z$ \\
\hline
3 & 1 & 3 & 0 & 0 & 0 & 0 \\
4 & -2 & -1 & 0 & 0 & 0 & 0 \\
5 & 1 & -1 & 0 & 0 & 0 & 0 \\
\hline \hline
\end{tabular}
\label{tab:InitialConditions}
\end{table*}

\begin{table*}[h]
\centering
\caption{Initial conditions for the D9 (3-body) problem. All
  parameters are presented in N-body units.  Scaling from
  dimensionless N-body units to physical units entails a size scale of
  9076\,au, and a mass-unit of $10^6$\,\MSun, leading to a
  time-scaling of $\sim 66$\,yr. Initial conditions are presented in
  16 decimal places: values of $0$ in input are equivalent to
  $0.0\rlap{0}/$.}
\begin{tabular}{
S[table-format=1.16]
S[table-format=1.16]
S[table-format=1.16]
S[table-format=1.16]
}
\hline \hline
\multicolumn{4}{l}{D9 and Sgr~A$\star$} \\
$m$ & $x$ & $y$ & $z$ \\
\hline
0.9999991784973042 & 0.0000005586218357 & 0 & 0 \\
0.0000006516168722 & -0.6800127100427656 & -0.0000032303170191 & 0.0000145110789701 \\
0.0000001698858274 & -0.6799485478701068 & 0.0000123902570596 & -0.0000556589330361 \\
\hline
& $v_x$ & $v_y$ & $v_z$ \\
\hline
& 0 & 0.0000011445677483 & 0 \\
& -0.0171540976955796 & -1.3899330802544423 & -0.0149447972483528 \\
& 0.0657965391063327 & -1.4060205340537189 & 0.0573225099936821 \\
\hline\hline
\end{tabular}
\label{tab:InitialConditionsD9}
\end{table*}

For the last experiment, we adopt a 100-body Plummer sphere. This is a
3-dimensional problem in the 7N-dimensional phase space. We selected
masses ranging over 2 orders of magnitude, following a power law with
an index of $-2.35$, representing a stellar mass function
\cite{1955ApJ...121..161S}. The initial system has a virial radius of
unity, a kinetic energy of 0.25, and a potential energy of -0.5 (all
these quantities are presented in dimensionless N-body units, i.e.,
Newton's constant $G=1$).  The initial system is virialized (with
virial ratio $Q=0.5$).

In the following section, we compare the identical initial conditions
when running with IEEE floating points, Posits and Universal of 16,
32, 64 bits, and with time-step parameters $\eta$ ranging from
$10^{-6}$ to $0.1$. 

\section{Results}\label{Sect:results}

Here, we present calculations using IEEE floats and Posits of the same
length. 16-bit word-length arithmetic was insufficient to obtain
astronomically meaningful calculations, irrespective of the
implementation.  For that reason we also perform a series of
experiments using 16-bit brain floating point (bfloat16)
\cite{Teich2018TPU3}.  We rely on the GCC (version 15.2.0 via the GCC
extension plus C++23 library support) supported brain-float via the
compiler built‑in type {\tt bf16} which exposes it as std::bfloat16\_t
in {\tt libstdc++}.  Our analysis, therefore, focuses on 32-bit,
64-bit, and in some cases, for validation purposes 128-bit quadruple
precision.

\subsection{Montgomery's figure-8 periodic 3-body solution}\label{Sect:Figure8}

The figure-8 braid is relatively easy to integrate numerically because
there are no close encounters and it is linearly stable.  We aim to
resolve 100 time units to measure performance and testing accuracy.
With an orbital period of $2\pi$ time units, we then resolve 15
orbits.  \cite{2000MNRAS.318L..61H} argue that the figure-8 problem
becomes dynamically unstable when changing one of the objects' masses
by $2^{-15} \sim 3 \cdot 10^{-5}$, which is comparable to the
precision reached with 16-bit arithmetic (${\cal O}(10^{-4})$).  With
a typical integration timestep is about $dt \simeq 2\pi/100$, 16-bit
floating point representation should then be sufficiently accurate to
resolve $>100$ orbits.

\begin{figure}[t]
\centering
\resizebox{\hsize}{!}
{\includegraphics[scale = 1.0]{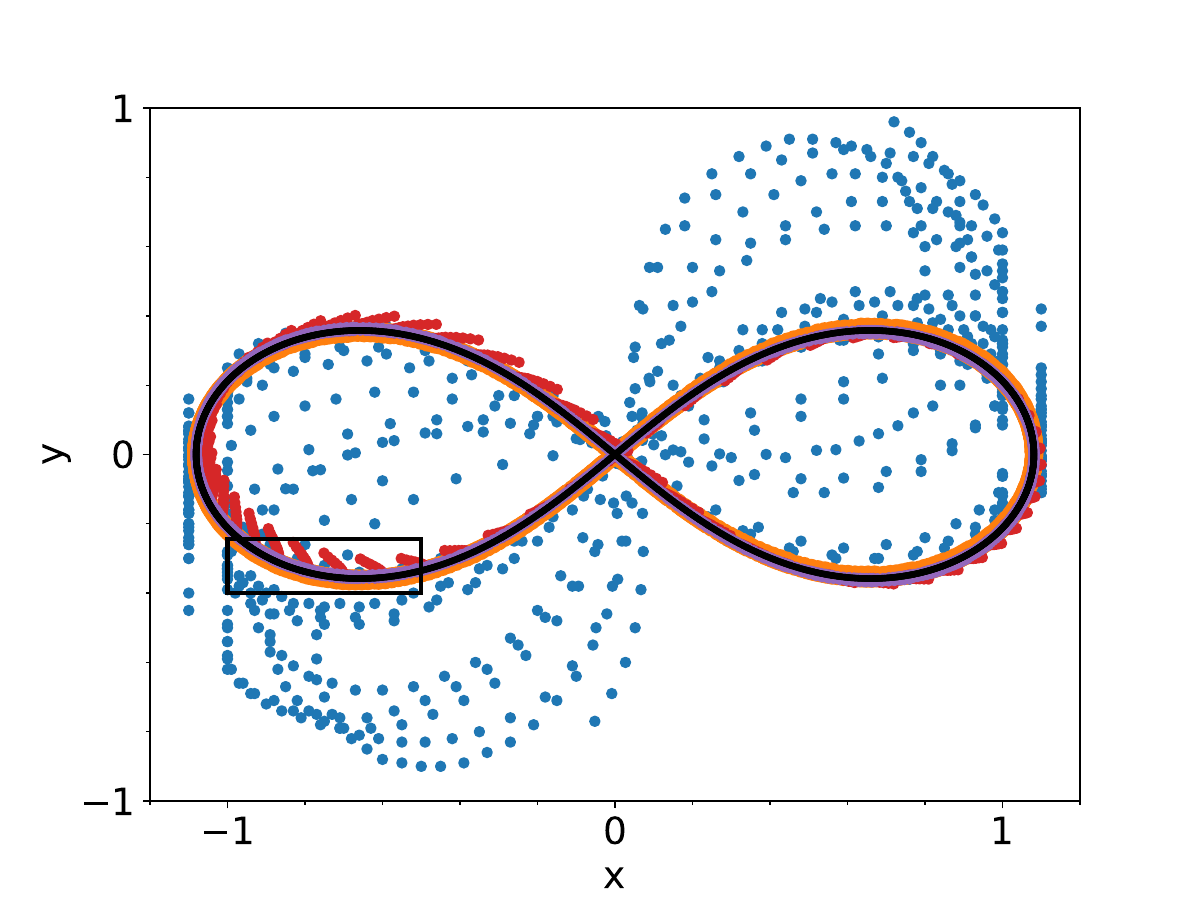}}
{\includegraphics[scale = 0.4]{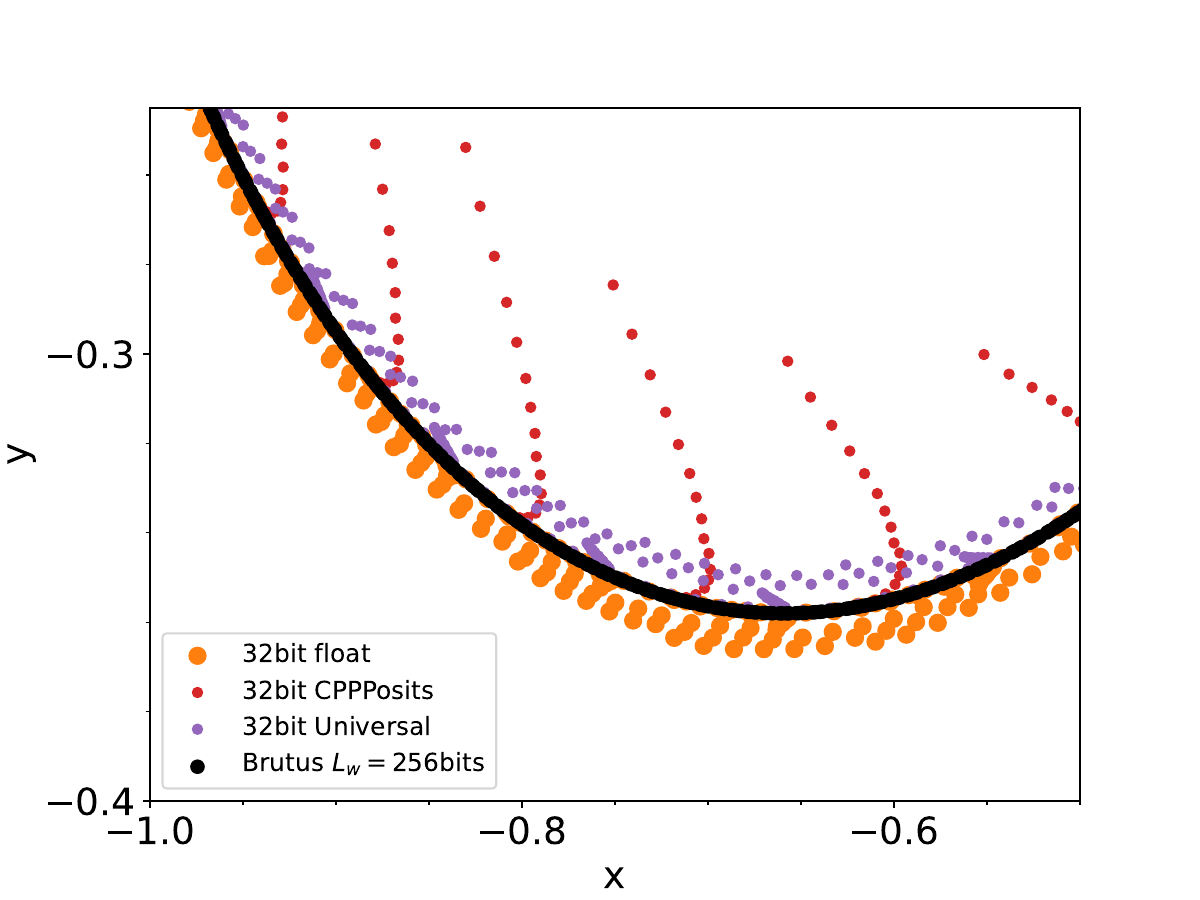}}
\caption{Orbit of figure-8's braid \cite{0951-7715-11-2-011}.
  Integration of the stable 3-body problem for 30 N-body time units
  (almost 5 orbits) using Brutus (converged with $L_w=256$\,bits) in
  black, IEEE single precision in orange, 32-bit CPPPosits in red,
  32-bit Universal in purple, and using 16-bit bfloat (blue, which we
  leave out of the bottom panel).  Calculations are performed with
  timestep $\eta = 10^{-4}$ (except for 16-bit bfloats for which $\eta
  = 0.1$). Diagnostic and data output time-step was 0.1 N-body time
  units. Both floats and CPPPosits deviate from the converged
  solutions, as further illustrated in the bottom panel, which shows a
  magnification of the highlighted area in the top panel. What look
  like spurs when following the red bullet points are the result of
  the precession of the orbit. }
\label{fig:Montgomery_Orbit}
\end{figure}

Still, 16-bit appears insufficient to complete even a single orbit,
irrespective of the time-step parameter $\eta$.  This improves
somewhat when adopting bfloat16 (half precision float with 8-bit
exponent rather than the usual 5 bits), which manages to complete two
orbits with $\eta = 0.1$, but fails for smaller values of $\eta$ (see
the blue dots in \cref{fig:Montgomery_Orbit}).
The bfloat16 outperforms float16 for the gravitational N-body problem
because of its larger dynamic range, allowing smaller time steps and
therefore more precise integraton.

In addition, we performed a 16-bit precision integration using
Brutus. Convergence for this calculation, was reached for $\epsilon =
10^{-8}$, basically forcing the Bulirsch-Stoer to converge to the
least significant digit. The resulting orbit has similar inappropriate
characteristics as those presented in \cref{fig:Montgomery_Orbit} for
16-bit bfloats (blue dots).

\begin{table*}[h]
\centering
\caption{Achieved precision at the run's end for the figure-8,
  Pythagorean, D9 and Plummer models.  The reference calculations were
  performed using Brutus with 256-bit precision with a tolerance
  $\epsilon=10^{-10}$. Except for the Plummer model reference run
  which was performed using $232$ bits and a Bulirsch-Stoer tolerance
  of $\epsilon = 10^{-50}$ Which is sufficient to reach convergence.
  The calculations using fp64, CPPPosits and Universal were performed
  in 64-bit precision.  The second column gives the adopted value of
  the time-step parameter $\eta$, followed with the time for which the
  data are compared. The following three columns give the number of
  precise digits averaged over all degrees of freedom (higher is more
  decimal places accurate).  The last three columns give the raw
  measurements on which the averages are based; they provide the same
  precision information but then with the number of free parameters
  for which the decimal places are preserved for fewer than 3
  decimals, 4/5/6 decimal places and 7 or more.  The dimensionality of
  the problem for the figure-8 and Pythagorean problems is 12, for D9
  it is 18-dimensional, and the $N=100$ Plummer model has 600 degrees
  of freedom. For the latter, we therefore present the fraction (in
  percent) of figures that achieve the indicated number of correct
  decimal places. For the $N=100$ Plummer model, we present a
  time-evolution of the calculation's significance in
  \cref{fig:Plummer_N100_precision}.  }
\begin{tabular}{p{2.0cm}p{0.6cm}R{1.0cm}|R{1.0cm}R{1.5cm}R{1.5cm}|R{2.0cm}R{2.0cm}R{2.0cm}}
\hline \hline
           & $\eta$   & t  & fp64    & CPPPosits  & Universal & fp64    & CPPPosits  & Universal \\
           &          &    & \multicolumn{3}{c|}{mean correct mantissa length} & ($<$3)3/4/5/6($>$6) & ($<$3)3/4/5/6($>$6) & ($<$3)3/4/5/6($>$6) \\
figure-8   & $10^{-4}$ & 100&  6.1 &  5.3 &  6.1& ~(0)~~0/0/1/9(2) & ~(0)0/0/9/3(0) & ~(0)~~0/0/1/9(2) \\
Pythagoras & $10^{-4}$ &  30&  4.4 &  4.6 &  4.3& ~(0)~~0/8/3/1(0) & ~(0)0/6/5/1(0) & ~(0)~~0/8/4/0(0) \\
D9         & 0.01     &1000&  4.4 &  4.4 &  4.9&  ~(4)~~1/3/4/0(4) & ~(5)2/4/1/0(6) & ~(2)~~3/1/4/2(5) \\
Plummer    & 0.01     &   6&  2.5 &  2.1 &  2.3& (73)13/9/4/1(0)& (92)5/3/0/0(0)& (80)13/5/2/0(0)\\
\hline \hline
\end{tabular}
\label{tab:FinalPrecision}
\end{table*}

In \cref{tab:FinalPrecision}, we present precision achieved in terms
of the number of parameters that reach a specific number of
significant digits for each of the 12 Cartesian coordinates.
Integrating figure-8's and the Pythagorean problems with Brutus
achieves 17 decimal places accurate for all the 12 degrees of freedom.
After 100-Nbody time units, this leads on average to $\sim 6$ correct
decimal places for Universal, and $\sim 5$ decimals for CPPPosits.  We
speculate that CPPPosits relatively bad performance, compared to fp64
and CPPPosits, results from differences in the implementation of the
quire (long accumulator); tabulated in CPPPosits and calculated in
Universal. This difference can also, in part at least, explain the
performance difference between the two implementations. However, if
the accumulator is the culprit, it remains a bit mysterious why the
fp64 calculations, which do not include the Kulisch accumulator,
reaches comparable precision as Universal.

The experiment with the widest dynamic range, the integration of the
binary D9, shows better precision for Universal than for fp64 (4.9
decimal places versus 4.4 for fp64, see \cref{tab:FinalPrecision}).
Interestingly, fp64 performs better in the 100-particle
Plummer model compared to Universal (2.5 decimal places on average for
fp64 versus 2.3 for Universal).  This latter may be caused by the
systematically different way in which Posits treat round-off compared
to fp64. Although both adopt the same algorithm, they can only round
to the nearest representable number, and the set of representable
numbers differs between the two formats.  The representable numbers
of fp64 are distributed uniformly on a logarithmic scale: each binade
contains $2^{52}$ numbers, irrespective of its magnitude. Posits have
tapered precision; a posit$\langle 64,3 \rangle$ has 58 fraction bits
for numbers near unity, but loses one fraction bit for each factor
$2^{8}$ further away from unity. Posit$\langle 64,3 \rangle$ therefore
has more fraction bits than fp64 for $2^{-48} \le |x| < 2^{48}$, and
fewer outside the range $2^{-56} \le |x| < 2^{56}$.  For
posit$\langle 32,2 \rangle$ the equivalent cross-over with fp32
occurs around $|x| \sim 2^{\pm 20}$.  In N-body units, the bulk of
the numbers in our calculations lie in the range where 64-bit Posits
are more densely spaced than fp64. The somewhat lower precision of
Universal for the Plummer model therefore does not result from a
coarser resolution of the posits, but from the different rounding
itself: different rounding leads to a different, though not
necessarily worse, round-off error in each operation. In a chaotic
system these differences are magnified exponentially, and the two
calculations eventually follow different trajectories, each of which
deviates from the converged solution in its own way.

In \cref{fig:Montgomery_Orbit}, we present the orbit for 30 N-body
time units using the converged solution by Brutus with 256 bits and a
tolerance of $10^{-10}$, IEEE-754 single precision, 32-bit CPPPosits,
and Universal.  For 32-bit precision, regular floats and Posits both
deviate from the converged solution (see \cref{fig:Montgomery_Orbit}).
We do not show the 64-bit Universal results because they appear
identical within the thickness of the lines; 64-bit Universal and fp64
both lead to comparable (and arguably reasonable) results (see
\cref{tab:FinalPrecision})).

\begin{figure}[t]
\centering
\resizebox{\hsize}{!}
{\includegraphics[scale = 1.0]{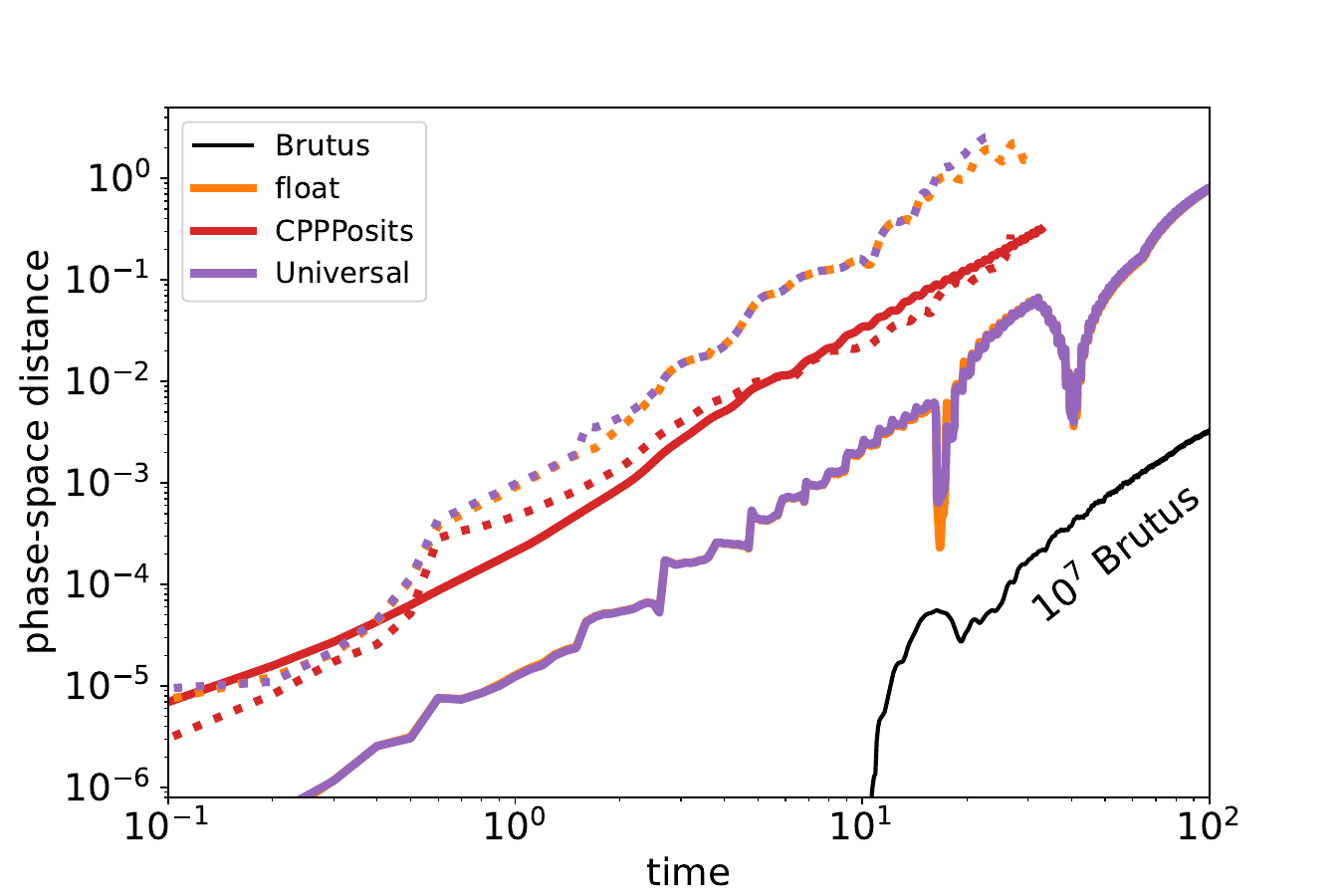}}
\caption{Phase-space distance evolution of several calculations using
  figure-8's initial conditions. Each curve gives the Cartesian
  distance between two solutions, one of which comes from the
  converged solution using Brutus with 256-bit precision. The other
  comes either from 4th order Hermite calculations using floats
  (orange), CPPPosits (red), or Universal (purple) using $\eta =
  10^{-4}$. The dotted curves (not shown for Brutus, as it results are
  Galileo invariant by construction) represent the same problem but
  displaced by 10 N-body units in each direction and with a velocity
  of 10 N-body velocity units in each direction. }
\label{fig:phsd_Mongomery}
\end{figure}

In \cref{fig:phsd_Mongomery}, we present the geometric distance
between a number of calculations for the figure-8 problem compared to
the converged reference solution with Brutus. The solid black line
compares the converged Brutus solution with a lower-precision solution
using Brutus with 64-bit word-length. For convenience, we multiplied
the phase-space distance by a factor $10^7$ to make the visual pattern
match better with the other solution. The solid set of curves compares
the converged Brutus results 32-bit floats (orange), CPPPosits (red),
and Universal (purple). In all cases, we adopted a time-step parameter
$\eta=10^{-4}$. Universal and fp64 produce a similar time evolution of
the phase-space distance, but for CPPPosits, the values are larger by
about an order of magnitude.  The slope, however, remains the same,
indicating that the system shows similar chaotic behavior (comparable
$t_\lambda$), but with an overall larger error in the numerics.  Chaos
in the gravitational N-body problem is not so much affected by
precision because it is a robust intrinsic physical quality of the
system.

The dotted curves result from the same calculation, but the initial
conditions (see \cref{tab:InitialConditions}) were displaced by 10
N-body units in any direction, and each particle was given a velocity
of 10 N-body velocity units in every direction ($v_x$, $v_y$, and
$v_z$). We perform this calculation to test Galilean
invariance\footnote{Galileo invariance is the notion that the laws of
physics are the same in any of the Cartesian coordinates.} of the
numerical representation.  It turns out that CPPPosits, Universal, and
fp64 suffer from under Galilean invarincy.

In terms of computing time, fp64 was the fastest, with $\sim 7 \cdot
10^{-5}/\eta$\,s for 100 N-body time units.  At the same precision
CPPPosits are a factor 22 slower than fp64, which is quite impressive
considering that the latter is hardware accelerated.  The Universal
implementation is roughly 1800 times slower than fp64.  Such
considerable loss of performance cannot be explained by a lack of
hardware support (see the discussion in \cref{Sect:Performance}).

\begin{figure}[t]
\centering
\resizebox{\hsize}{!}
{\includegraphics[scale = 1.0]{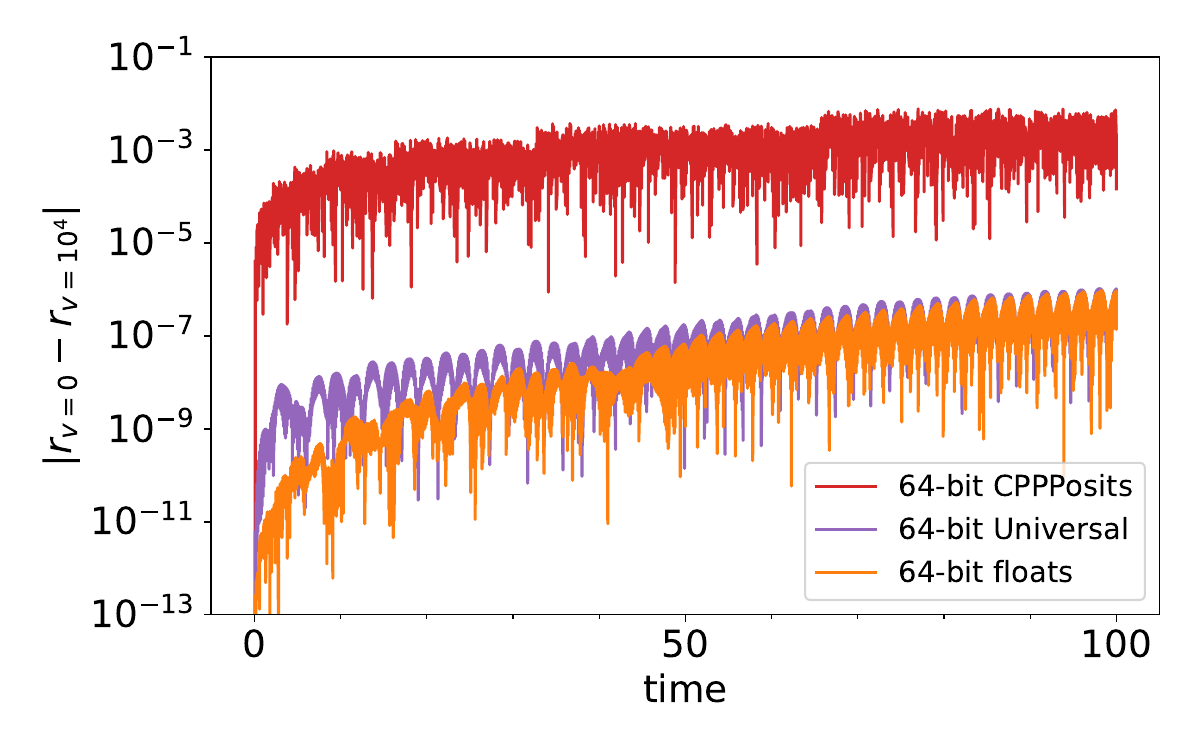}}
\caption{Phase space distance between two solutions of the figure-8
  problem for fp64 (orange), 64-bit CPPPosits (red), and Universal (purple). One of the solution is
  and remains in the center of 3-body mass, the other solution
  moves with a velocity of $10^4$ N-body speed units in the $x$, $y$,
  and $z$ direction (see also \cref{fig:Pythagorean_Galileo}). 
  \label{fig:Montgomery_Galileo}
  }
\end{figure}

In \cref{fig:Montgomery_Galileo}, we show the phase space distance
between two solutions of the figure-8 problem for fp64, CPPPosits, and
Universal.  The moving system quickly starts producing relatively
large deviations, eventually losing 7 decimal places for fp64, and
Universal, and even more for CPPPosits. In particular in the beginning
of fp64 performs better than Universal, but this different becomes
negligible after about 5- N-body time units, when both deviations
saturate around $10^{-7}$. CPPPosits make larger Galilean invariance
errors, on the order of $10^{-3}$.

The large effect of moving the system in all three Cartesian
coordinates should probably not comes as a surprise, but we were still
somewhat struck by the impact of introducing a systemic velocity to
the system. Note, that, as expected, the error introduced by moving
the system is proportional to the velocity. We tested velocities from
0 to $10^8$, causing the resulting error to gradually degraded (see
\cref{fig:Montgomery_Galileo}).

\subsection{Pythegorean}\label{Sect:Results.Pythagorean}

The Pythagorean 3-body problem is considerably more complex to
integrate than the figure-8 problem. Here, close encounters tend to
drive numerical errors, even though the interaction lasts for only 60
N-body time units, few integrators manage to reach a converged
solution. A converged solution to 40 decimal places after 100 N-body
time units was presented in \cite{2018CNSNS..61..160P}. Their solution
was achieved with wordlength $L_w=136$-bit precision and a tolerance of
$\epsilon=10^{-24}$. In our case, neither of the implementations
resolves this problem even to a single accurate digit.

\begin{figure}[t]
\centering \resizebox{\hsize}{!}  {\includegraphics[scale =
    1.0]{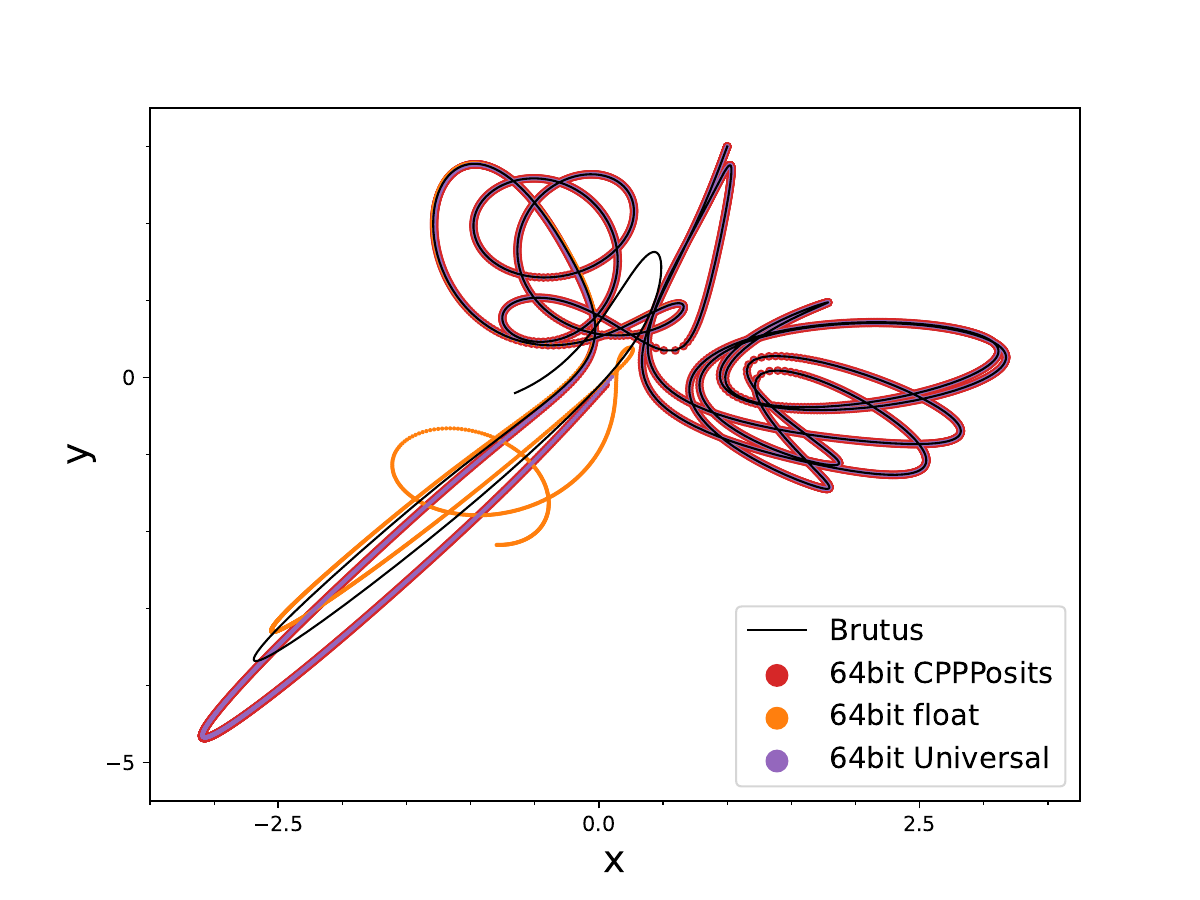}}
\caption{Orbit of the top corner particle (mass=3) in the Pythagorean
  3-body problem for 60 N-body time units. Diagnostic and data output
  time-step were 0.01 N-body time units using a time-step parameter of
  $\eta = 10^{-2}$. }
\label{fig:phsd_Pythagorean_orbit}
\end{figure}

In \cref{tab:FinalPrecision}, we present precision achieved for the
Pythagorean problem's 12 degrees of freedom after 30 N-body time
units, using a time-step parameter $\eta= 10^{-4}$.  As we saw in
\cref{Sect:Figure8}, fp64 and Universal reach comparable precision
($\sim 4.4$ significant digits), whereas Posits ($\sim 4.6$
significant digits) perform even slightly better.  This is easiest
seen by the lower mean number of accurate decimal places for fp64 and
Universal compared to CPPPosits in \cref{tab:FinalPrecision}.  It may
seem curious that CPPPosits outperform fp64 and Universal here, but by
the end of the simulation both systems have deviated considerably and
they have become incomparable (see
\cref{fig:phsd_Pythagorean_orbit}).

In \cref{fig:phsd_Pythagorean_orbit}, we present the orbit of one of
the three stars in the Pythagorean problem. Here, black gives the
converged solution, orange the pf64, red represents 64-bit CPPPosits, and
purple the solution calculated with the Universal library. At $t=0.01$
N-body time units, the non-converged solutions already deviate from
the converged solution (and each other) in the 4th decimal place. By
$t=0.1$ N-body time units, both solutions have deviated to the second
decimal place.

Still, all thee representations remain relatively close to the
converged solution.  The solutions start to deviate visibly near
$t\sim 25$, around the large loop to the lower left corner in
\cref{fig:phsd_Pythagorean_orbit}. This loop is initiated by a
relatively close encounter.  The fp64 solution remains closest to the
converged solution, but CPPPosits and Universal perform reasonably
well.  In this case, the strong deviation from the nominal solution
originates from a numerical error during a close encounter, rather
than from the system's chaotic dynamical evolution. By the end of the
run (at $t=60$ N-body time units), all three solutions have deviated
from the converged solution even to the first decimal place.  Since
the system is chaotic, the solutions cannot be compared any more and
it would go too far to interpret the curious ``pig tail''
behavor of fp64 near the simulation's end.

\begin{figure}[t]
\centering
\resizebox{\hsize}{!}
{\includegraphics[scale = 1.0]{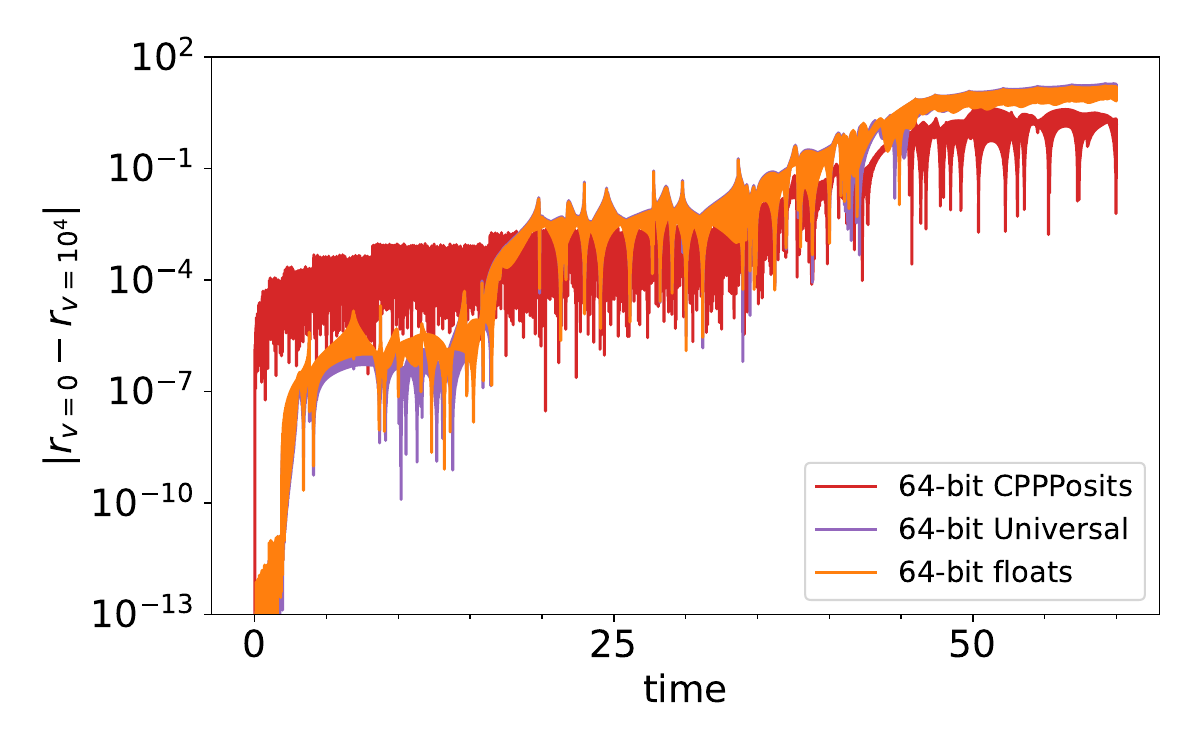}}
\caption{Phase space distance between two solutions of the Pythagorean
  problem for fp64 (orange), 64-bit CPPPosits (red), and Universal
  (purple). One of the solution is and remains in the center of
  3-body mass, the other solution moves with a velocity of $10^4$
  N-body speed units in the $x$, $y$, and $z$ direction (see also
  \cref{fig:Montgomery_Galileo}). }
\label{fig:Pythagorean_Galileo}
\end{figure}

\Cref{fig:Pythagorean_Galileo} shows the Pythagorean equivalent for
\cref{fig:Montgomery_Galileo}. In this case, however, Universal
performs comparably as fp64, whereas CPPPosits produce errors several
orders of magnitude above those of fp64. At a later time, after about
$t=25$ CPPPosits perform better than the other two, which seems
somewhat surprising. However, by that time, the orbital integrations
of CPPPosits as well as fp64, and Universal have deviated that we
cannot directly compare the results (as can be see in
\cref{fig:phsd_Pythagorean_orbit}).  Possibly, Universal and fp64
remain relatively strongly bound, whereas the CPPPosits solution
softens.

\subsubsection{Energy conservation}

So far, we have not discussed the energy error of the simulations as a
function of time-step parameter $\eta$. The parameter $\eta$ gives a
linear relation between the minimum free-fall time-scale in the
gravitational N-body problem and the actual time step taken by the
integrator.

Usually, the simulation time increases inversely proportional to the
value of $\eta$. Simulations with small values of $\eta$ then become
expensive, in particular if those are performed without hardware
support. We therefore stop at $\eta = 10^{-6}$.

\begin{figure}[t]
\centering
\resizebox{\hsize}{!}
{\includegraphics[scale = 1.0]{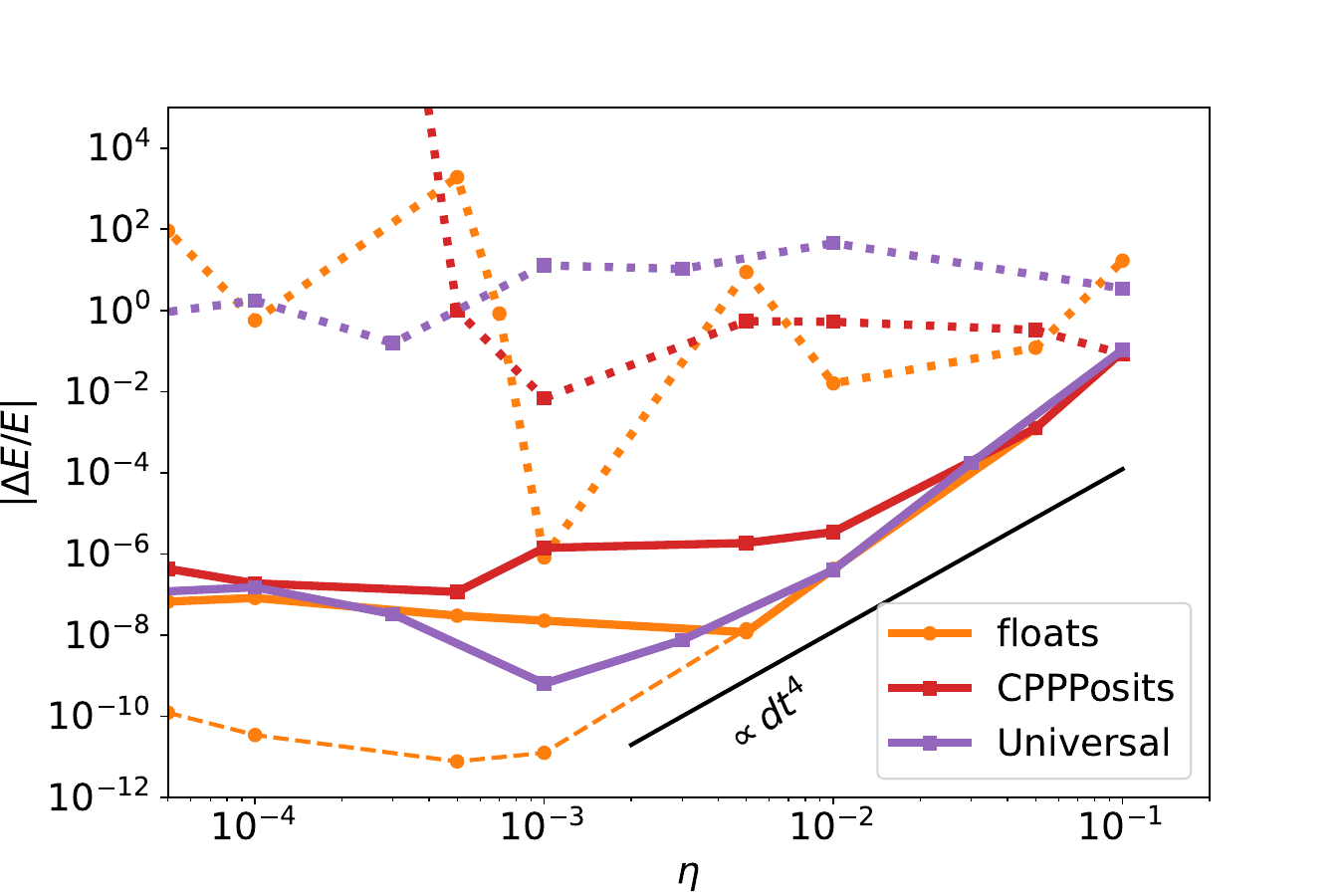}}
\caption{Relative energy error of the integration as a function of the
  time-step parameter ($\eta$) for the Pythagorean problem integrated
  over 60 N-body time units. The calculations are for floats (orange)
  CPPPosits (red), and Universal (purple), for 32-bit (dotted lines),
  64-bit (solid), and 128-bit (orange dashed line). }
\label{fig:dEoverE_Pythagoras}
\end{figure}

In \cref{fig:dEoverE_Pythagoras}, we present the relative difference
in the energy of the Pythagorean 3-body system as a function of the
time-step parameter $\eta$. The minimum energy error for fp64 (orange
curves) is reached for values of $\eta \simeq 5 \cdot
10^{-3}$. Smaller values of $\eta$ lead to larger energy errors due to
the increased effect of round-off, whereas for values of $\eta >
10^{-3}$, larger errors are due to large time steps. The latter scales
roughly with $\Delta E/E \propto dt^{4}$ because of the
characteristics of the adopted 4-th order integration algorithm.

For 64-bit CPPPosits (red solid curve) the energy error saturates
already around $\eta \simeq 10^{-2}$. This trend confirms our earlier
suspicion that CPPPosits occasionally produces unrecoverable errors
while integrating; Here Universal performs slightly better than fp64.

None of the 32-bit solutions (dotted curves in
\cref{fig:dEoverE_Pythagoras}) is sufficiently reliable for any
scientific computation; these results are dominated by round-off
errors over the entire range of $\eta$. They systematically fail to
keep the relative energy error below $10^{-2}$ for any value of
$\eta$. According to \cite{2018CNSNS..61..160P}, such solutions are
apprehensive (case B) and cannot be reliably used for scientific
interpretation.

\subsection{The binary D9 orbiting the Galactic central black hole}\label{Sect:Results.D9}

To further test the application range for CPPPosits and Universal, we
adopt the recently discovered binary star, D9, in orbit around the
supermassive black hole in the Galactic center (a.k.a.\, Sgr
A$\star$). Orbital integration of D9 is less prone to errors in energy
and angular momentum. The challenge in this 3-dimensional problem
hides in its secular stability, and its wide dynamic range, both of
which can be challenging for any numerical representation.

The binary D9, discovered by \cite{2024NatCo..1510608P}, is composed
of two stars of $2.8\pm0.5$\,\MSun, and a $0.7 \pm 0.1$\,\MSun\, in a
$1.59\pm 0.01$\,au orbit with an eccentricity of $e=0.45$. The binary
orbits the black hole in the Galactic center (Sgt\,A$\star$) at a
distance of 0.44\,mpc (about $\sim 9100$\,au) with an eccentricity of
$\sim 0.32$. The velocity of D9 at peribothron $\sim 380$\,km/s, and
the maximal relative orbital velocity of the two stars in D9 exceeds
$250$\,km/s.

The black hole (Sgt\,A$\star$ has a mass of about
$4.3\cdot10^6$\,\MSun\, \cite{1997MNRAS.284..576E,1998ApJ...509..678G}
but for our calculations we adopted $10^6$\,\MSun). Interestingly
enough, the orbit of D9 is inclined with respect to its orbit around
the black hole by about $103\pm 2^\circ$. The tight binary, D9, is
therefore subject to von Zeipel-Lidov-Kozai (vZLK) cycles
\cite{1962AJ.....67..591K}; a periodic exchange of angular momentum
between the inner orbit (of D9) and the outer orbit (D9's orbit around
the black hole). These cycles are a secular non-chaotic dynamical
process, which are challenging to reproduce through numerical
integration. With these parameters the vZLK cycle takes about
60-thousand years.

Integrating this system for $t=2000$\,N-body time units using Brutus
with 256-bit precision converges to 65 significant digits, to 29
significant digits when run with 128-bit IEEE floating point
precision, and to 11 significant figures when adopting 64-bit
precision. In \cref{tab:FinalPrecision}, we compare the precision
reached with the various numerical representations for D9's 18 degrees
of freedom when integrated for 1000 time units with $\eta = 0.01$.

\begin{figure}[t]
\centering
\includegraphics[scale = 0.5]{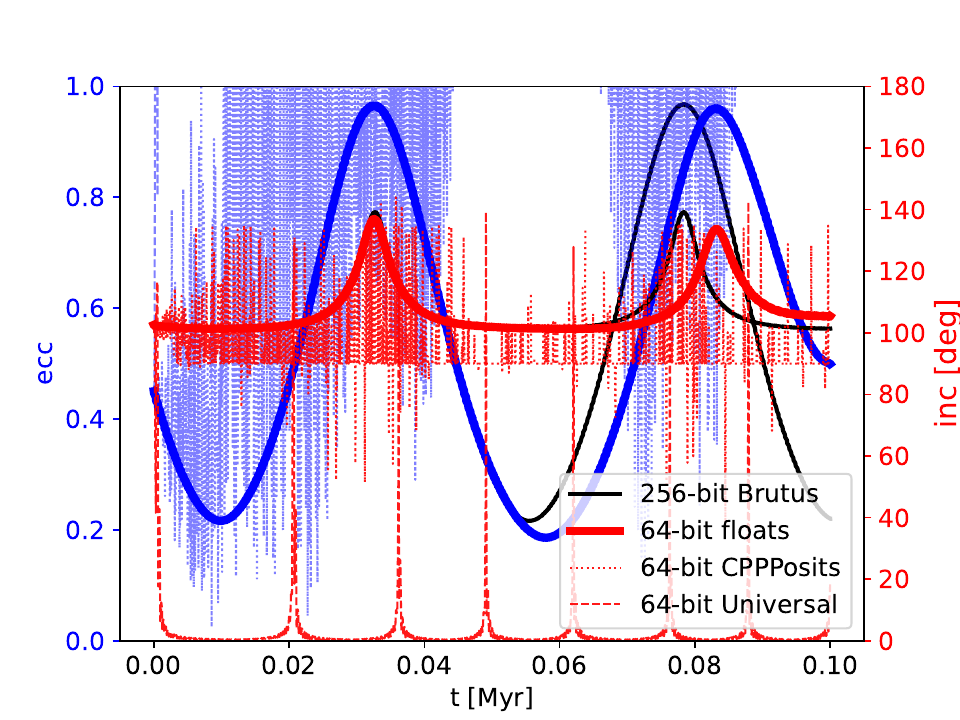}
\caption{Galileo invariant calculation for D9. In this experiment D9
  and the black hole was given a starting position offset by 10\,au,
  and a velocity of 10km/s in each Cartesian direction. We present the
  eccentricity (blue) and inclination (red) for the binary D9 as a
  function of time. The solid curves give the results for 64-bit
  floats, and 64-bit CPPPosits are given by the dotted curves. The
  latter does not seem to engular momentum appropriately. The fp64 and
  Universal (not shown) perform better, and stay close to the results
  calculated with Brutus. The black solid curve gives the calculation
  using Brutus (with 256-bit word-length): The sinusoidal curve
  represents the eccentricity evolution, and the more pointed curve
  gives the inclination. We omit the Universal solution because it is
  indistinguishable from the fp64 results.  }
\label{fig:D9_offset}
\end{figure}

To test the numerical stability of Galileo invariance, we offset the
entire system by 10\,au in each Cartesian coordinate and give it a
velocity of 10\,km/s in each direction. The results of this
calculation are presented in \cref{fig:D9_offset}, with Brutus (thick
curves), fp64 (thin curves), and CPPPosits (dotted curves).

Again, Brutus provides a baseline ground-truth solution. While fp64
and Universal give indistinguishable results compared to the 64-bit
non-Galilean invariant experiment. Posits fail in resolving the orbit
of D9 around the black hole. The two stars continue to orbit the black
hole, but their mutual orbit is broken up. 


\begin{figure}[t]
\centering \includegraphics[scale = 0.3]{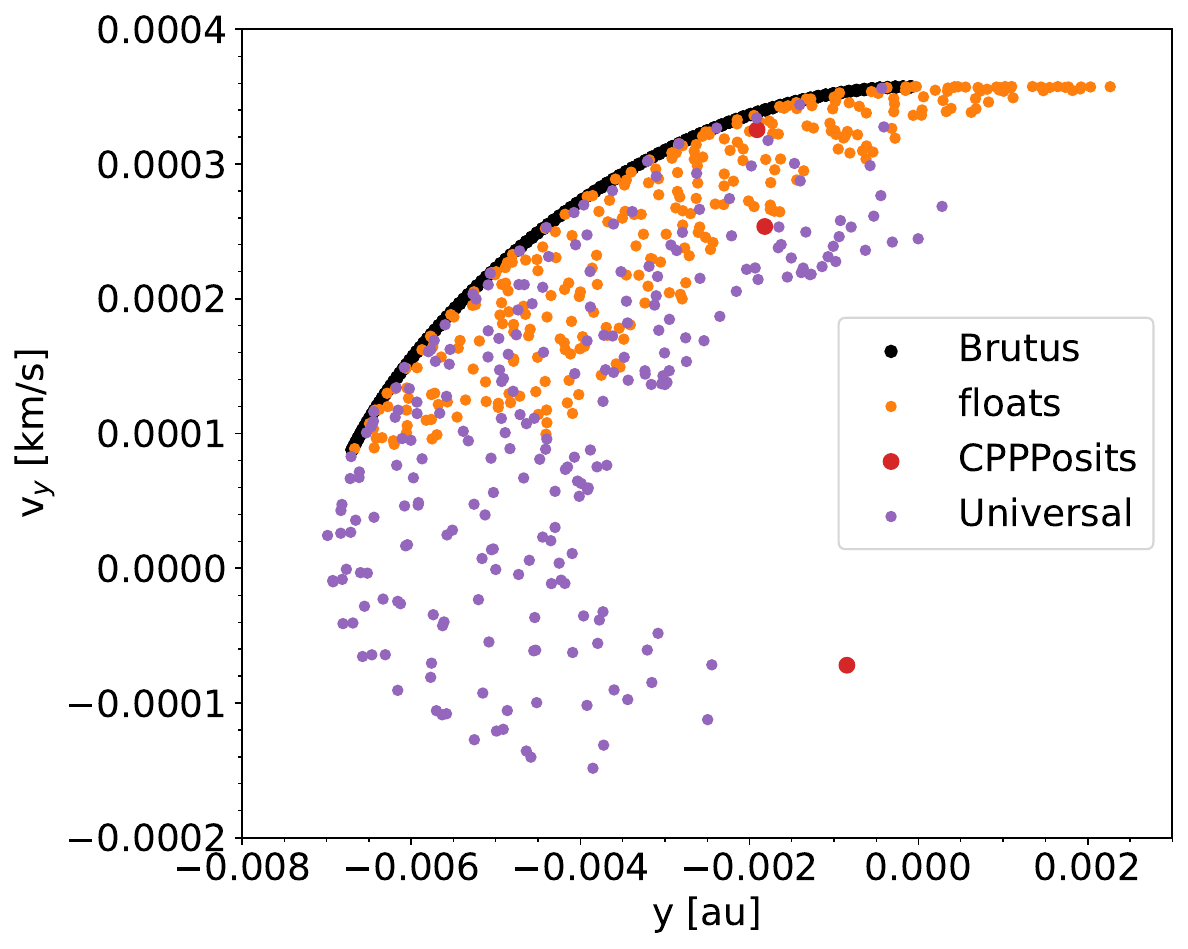}
\caption{Poincar{\'e} section of the black hole's orbit while
  integrating D9. The converged Brutus solution (using 256-bit
  precision) is represented with the black bullets. The orange, red,
  and purple bullets give the results calculated with 32-bit float,
  Posits, and Universal, respectively. The coordinates calculated using Posits are
  clustered in proximity in three distinct points.}
\label{fig:D9_Poincare}
\end{figure}

In \cref{fig:D9_Poincare}, we present Poincar{\'e} sections for the
black hole (Sgr A$\star$) when it passes through the Cartesian
x-coordinate with positive velocity. Being $\sim 10^6$ times more
massive than the binary D9 orbits, the black hole is near the system's
center of mass (see also \cref{tab:FinalPrecision}). Resolving the
black hole orbit is essential for properly integrating the binary D9,
and hard to resolve accurately. The black hole is located only about
0.005\,au from the system's center of mass, whereas the binary orbits
at a distance of almost 9100\,au, requiring already more than 6
decimal places to resolve. A similar problem arises in resolving the
mass of the two stellar components of D9 compared to the black hole
mass. As a consequence, 16-bit precision lacks dynamic range, and
32-bit precision remains questionable; All 16-bit and 32-bit
calculations fail to resolve the subtle triple dynamics.

The black hole's orbit is not chaotic on the time frame of the
integration. This is demonstrated in \cref{fig:D9_Poincare} where the
converged solution, calculated using Brutus and presented as the black
dots, follows a smooth curve. The solution was calculated with 32-bit
floats and Universal give rather scattered views, indicating chaotic
motion in both numerical representations.

The origin of chaos in this solution is caused by the numerical
round-off of the least significant digit. The fact that the
distribution of the point in fp64 and Universal do not overlap is
harder to interpret, but it indicates that both numerical
representations drive chaos in the system differently.  This is not
entirely surprising, as the rounding error driving chaos in these
cases differs between fp64 and Universal formats. Both round to the
nearest appropriate value, but these values are distributed
differently across their respective allotted ranges.

Posits with 32-bit precision overlay on three locations along $x \sim
-0.001$, indicating the loss of numerically driven chaos, but also the
inability of the numerical precision to properly represent the black
hole's motion. The resulting orbit of D9, being dominated by the black
hole, cannot be resolved. From this experiment, we conclude that all
three numerical representations fail at 32-bit precision, but recover
the converged orbit when adopting 64-bit precision.

Note that the numerical behavior of the system is peculiar. Where
arbitrary precision leads to a regular orbit, limited (32-bit floats)
precision exhibits chaos. So far, I have not seen a clearer example of
numerically-driven chaos. When we change numerical representation
floats to Posits, the solution regularizes. This behavior roots in the
differeces in treating the quire and the round-off in Posits (both
implenentations) compared to fp64.

\subsection{Plummer $N=100$}\label{Sect:N100Plummer}

For the final and hardest problem, we adopt 100 point-mass particles
distributed in a virialized Plummer \cite{1911MNRAS..71..460P} sphere
as initial conditions. This problem is exponentially sensitive to
infinitesimal perturbations to the initial conditions and to runtime
errors. As a consequence, small time steps and many decimal places are
required to reach a converged solution, making running this problem
expensive in terms of computer time.

A converged solution is reached after iteratively reducing the
tolerance $\epsilon$ while increasing the word-length $L_w$, starting
with $\epsilon = 10^{-8}$ and $L_w=16$. A converged solution for $n=5$
decimal places to $t=6$ N-body time units was obtained by running
Brutus with Bulirsch-Stoer tolerance $\epsilon = 10^{-50}$ and a word
length of $L_w=232$ (a mantissa of 58 digits).  We did not pursue the
calculation to $L_w=256$, as the results with Brutus, which are very
expensive in terms of computer time, already converged at the
$L_w=232$; Adopting a longer mantissa would not have changed the
results.  Shortly after this a particularly tight binary forms that
dramatically reduces the system's Lyapunov time scale. As a
consequence, the distribution takes a rather steep dive.  It took
about two weeks on 8 cores of a Xeon-based workstation to complete the
calculation.

\subsubsection{Quantitative analysis for $N=100$ Plummer}


\begin{figure}[t]
\centering
\resizebox{\hsize}{!}
{\includegraphics[scale = 1.0]{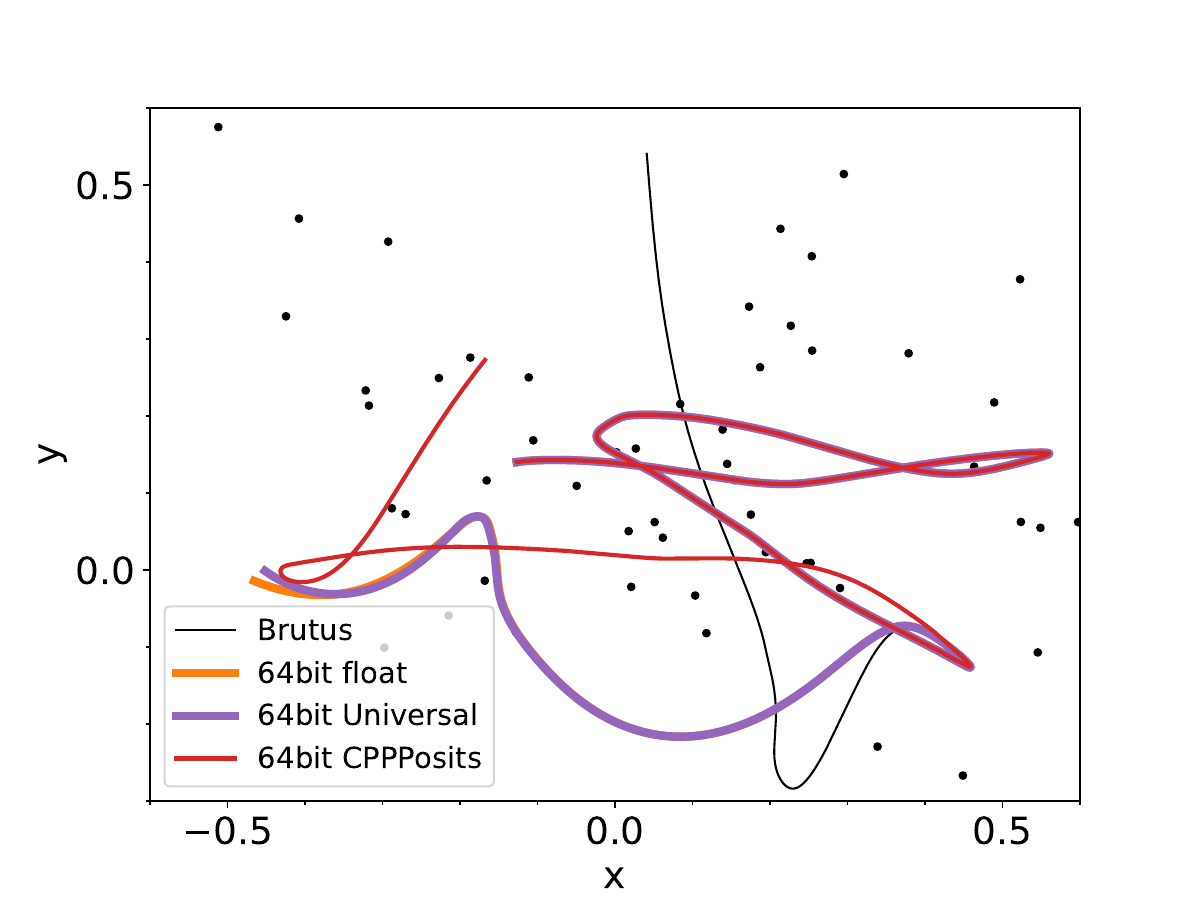}}
\caption{Orbit of one particle (number 6 with a mass of $\sim 0.00044$
  somewhat less than the mean of $0.00056$) of the 100 particle
  Plummer sphere (the initial conditions are represented with the
  black bullet points). The black curve gives the converged (to 4
  decimal places) Brutus solution. The diagnostic and data output
  time-step was 0.0001 N-body time units. IEEE-754 double precision in
  orange, 64-bit CPPPosits in red, and Universal in purple. }
\label{fig:PlN100_orbit_id6}
\end{figure}

\begin{figure}[t]
\centering
\resizebox{\hsize}{!}
{\includegraphics[scale = 1.0]{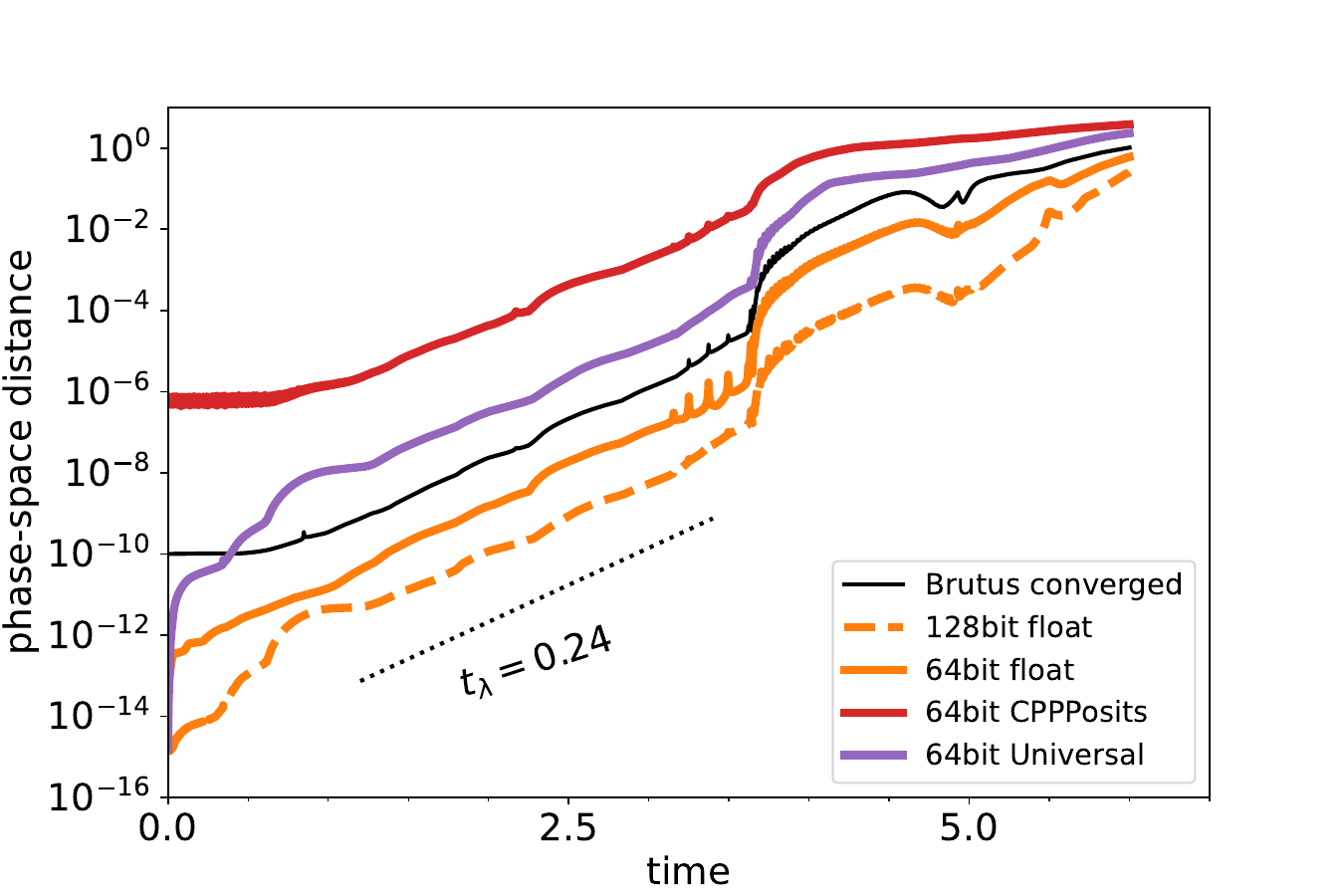}}
\caption{Phase-space distance between the various solutions of the
  same initial realization for 100 equal-mass particles in a
  virialized Plummer sphere. The three curves (red, orange, and
  purple) have similar shapes and slopes but are offset by a
  considerable margin. The black dotted line gives the growth rate,
  and its reciprocal relates to a Lyapunov time scale of $t_\lambda
  \simeq 4.17$. }
\label{fig:phsd_PlN100}
\end{figure}

In \cref{fig:PlN100_orbit_id6}, we present the orbit of one of the
stars in our sample for $t=6$\, N-body time units. The calculation
with fp64 and Universal manages to stay close to the converged orbit
for longer than 64-bit CPPPosit. The selected star had a relatively
close encounter around $t\sim 3.6$ (about halfway through the run),
causing all the 64-bit solutions to make an unrecoverable error,
driving their solutions away from the converged solution. This event
is also noticeable as the sudden increase in the phase-space distance
in \cref{fig:phsd_PlN100}.

The deviation of the four curves around $x=0.35, y=-0.1$ in
\cref{fig:PlN100_orbit_id6} indicate a resolution problem, most
prominently visible in CPPPosits but also present among fp64 and
Universal. After this close encounter, both the fp64, and Universal
solutions continue to shadow each other until they start to deviate
near the end (to the left of \cref{fig:PlN100_orbit_id6} near $y=0$).

We perform an additional simulation using 128-bit quadruple precision
IEEE-754 arithmetic. The 128-bit solution (not presented in
\cref{fig:PlN100_orbit_id6}) manages to resolve the close encounter
and continue to shadow the converged solution, presented as the black
thin curve (to within a line thickness).

In \cref{fig:phsd_PlN100}, we present the phase-space distance of the
various representations compared to the converged solution calculated
with Brutus. The 128-bit (dashed orange curve) and 64-bit (solid
orange) floats show similar dynamical behavior but at a systematic
offset of three orders of magnitude. The calculations performed with
Posits (64-bit CPPPosits and Universal, red and purple in
\cref{fig:phsd_PlN100}, respectively) deviate from the two IEEE
solutions (fp64 and fp128). Most noticeable is their failure to
resolve the rapid variation in the phase-space distance between
$t=3.2$ and $t=3.5$. The underlying inability of CPPPosits to
consistenly resolve close encounters leads to occasional large
rounding erros, whereas fp64 and Universal seem to provide more robust
results.

\subsubsection{Qualitative analysis for $N=100$ Plummer}

In \cref{fig:phsd_PlN100}, we also present the phase-space distance
between two converged calculations (black curve, both performed with
Brutus with the same parameter settings). The second calculation,
called Br$^p$, is identical to the first, but the x-coordinate of the
first particle in the initial realization was displaced by $10^{-10}$.
This infinitesimal perturbation between the two runs grows
exponentially, much in the same way as numerical round-off grows
exponentially. In this experiment, however, round-off remains well
below the lower y-axis boundary. Both solutions are converged, and the
growth of the perturbation gives the actual phase-space growth of the
chaotic system. This curve, therefore, representing an intrinsic
quality of the system rather than the consequence of a numerical
representation. The vertical offset of $\sim 10^{-10}$ gives rise to
the starting location of the phase-space distance curve in
\cref{fig:phsd_PlN100}. The phase-space distance between these two
solutions closely follows the fp64 calculation.

\begin{figure}[t]
\centering
\resizebox{\hsize}{!}
{\includegraphics[scale = 1.0]{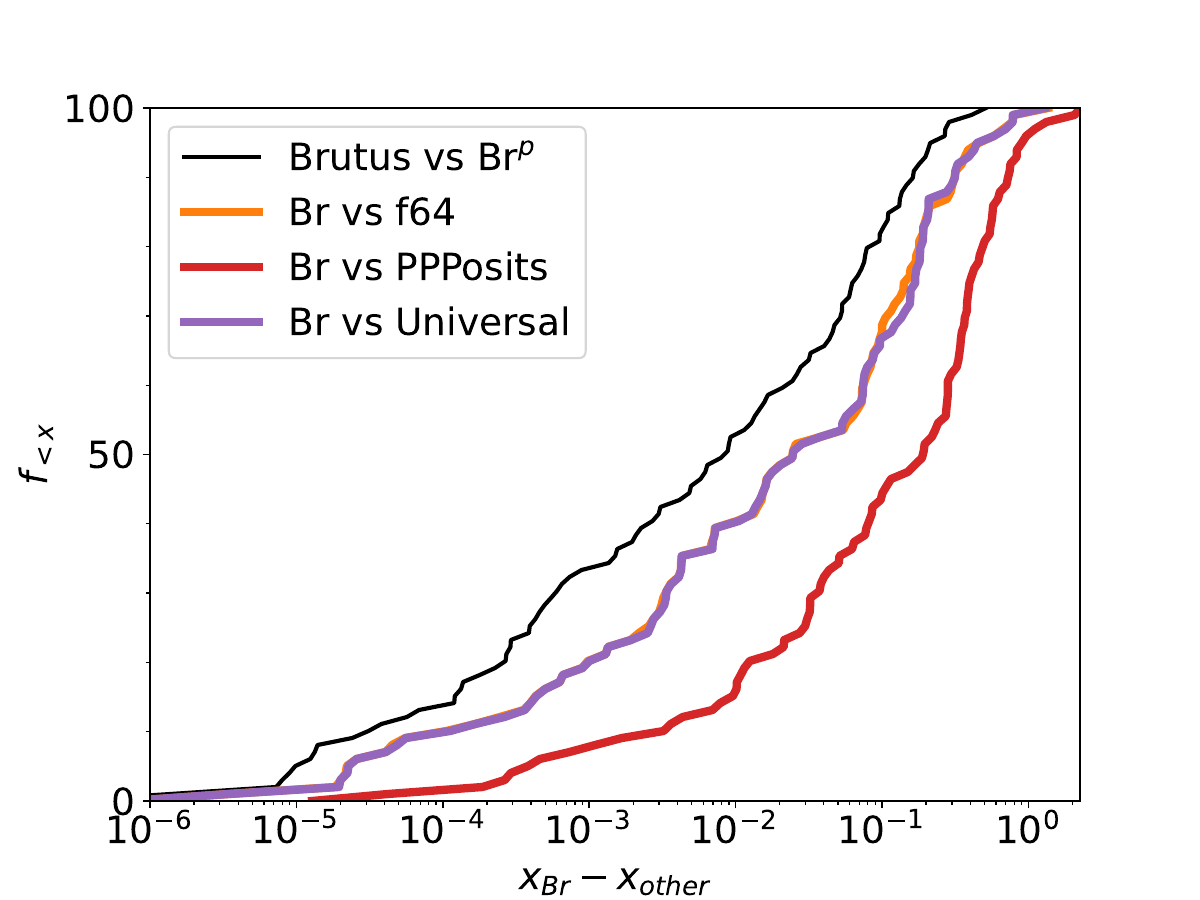}}
\caption{Comparison between final snapshots calculated with Brutus
  (Br) and another arithmetic representation.  The black curve
  compares Brutus with the perturbed Brutus solution.  Orange compares
  with fp64, red with CPPPosits, and purple with Universal.}
\label{fig:PlN100_hist}
\end{figure}

To further quantify the differences in the various numerical
representations we present, in \cref{fig:PlN100_hist} the cumulative
distributions of the relative phase-space distance between 64-bit
CPPPosits (red), Universal (purple), fp64 (orange), and converged simulations using Brutus (black):
\begin{equation}
\Delta \equiv \log_{10}(d_{\rm 64bit}/d_{\rm Brutus}) = \log_{10}(({\rm Br-64bit})/({\rm Br-Br^p})).
\label{Eq:phase_space_distance}\end{equation}
Here, Br represents the positions of the 100 particles at the age of 6
N-body time units calculated with the converged Brutus calculations.
Br$^p$ gives, for the same particles, the position at the same
time. Brutus solutions were converged to 16 decimal places. As a
consequence, the Br and Br$^p$ simulations give the actual phase-space
distance evolution of the initial 100-body Plummer model, and could
therefore be considered an estimate for $t_{\lambda}$.  Both the
64-bit solutions are calculated for IEEE-754 double-precision, and for
64-bit CPPPosits. In \cref{fig:PlN100_hist}, we present the cumulative
distribution of the ratio of the relative distance between the
converged (unperturbed minus perturbed) solutions with the converged
(unperturbed) minus the specific floating-point representation.

The distribution of errors generated in fp64 and Universal, presented
in \cref{fig:PlN100_hist}, have the same mean values of $\langle
\Delta_{\rm floats} \rangle = 0.50 \pm 0.34$.  Whereas for Posits we
find $\langle \Delta_{\rm Posits} \rangle = 1.27 \pm 0.64$.  64-bit
CPPPosits results in a considerably broader distribution at an, on
average, larger value.  The converged Brutus results give, not
surprisingly, the narrowest distribution. CPPPosits, for this specific
experiment, produce less reliable, less predictable results because
they lead to larger errors (on average), and the errors exhibit a
wider range.

\begin{figure}[t]
\centering
\resizebox{\hsize}{!}
{\includegraphics[trim=0cm 0cm 33cm 0cm, width=1\textwidth]{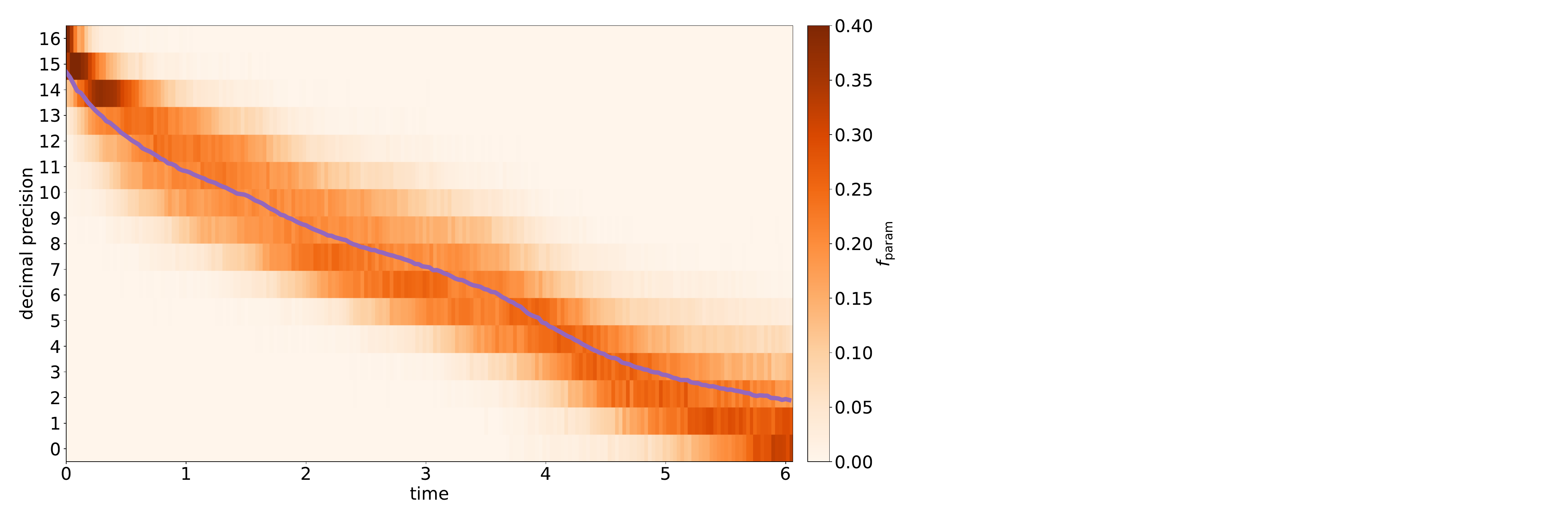}}
\caption{Evolution of the distribution of the number of decimal places
  that represents the parameter space for the $N=100$ Plummer model.
  We express this in $f_{\rm param}$, or the fraction of parameters
  that at time $t$ has number of decimal places accurate.  The orange
  shades give the result for fp64, the purple curve gives the mean
  number of accurate positions for Universal. Calculations were
  performed for $\eta = 0.01$. The parameter space is $6N$ dimensional
  (three coordinates in Cartesian position and three in velocity for
  each particle).  Time along the x-axis, and along the y-axis we find
  the number of decimal places of the numerical precision. The colors
  represent the fraction of the parameter space that is represented
  with that specific precision.  }
\label{fig:Plummer_N100_precision}
\end{figure}

In \cref{fig:Plummer_N100_precision}, we present the results of the
$N=100$ Plummer model, comparing the fraction of decimal places to
which the 600 parameters are correctly represented in the fp64
solution (masses are assumed to remain constant).

\begin{figure*}
\centering
\includegraphics[width=1\columnwidth]{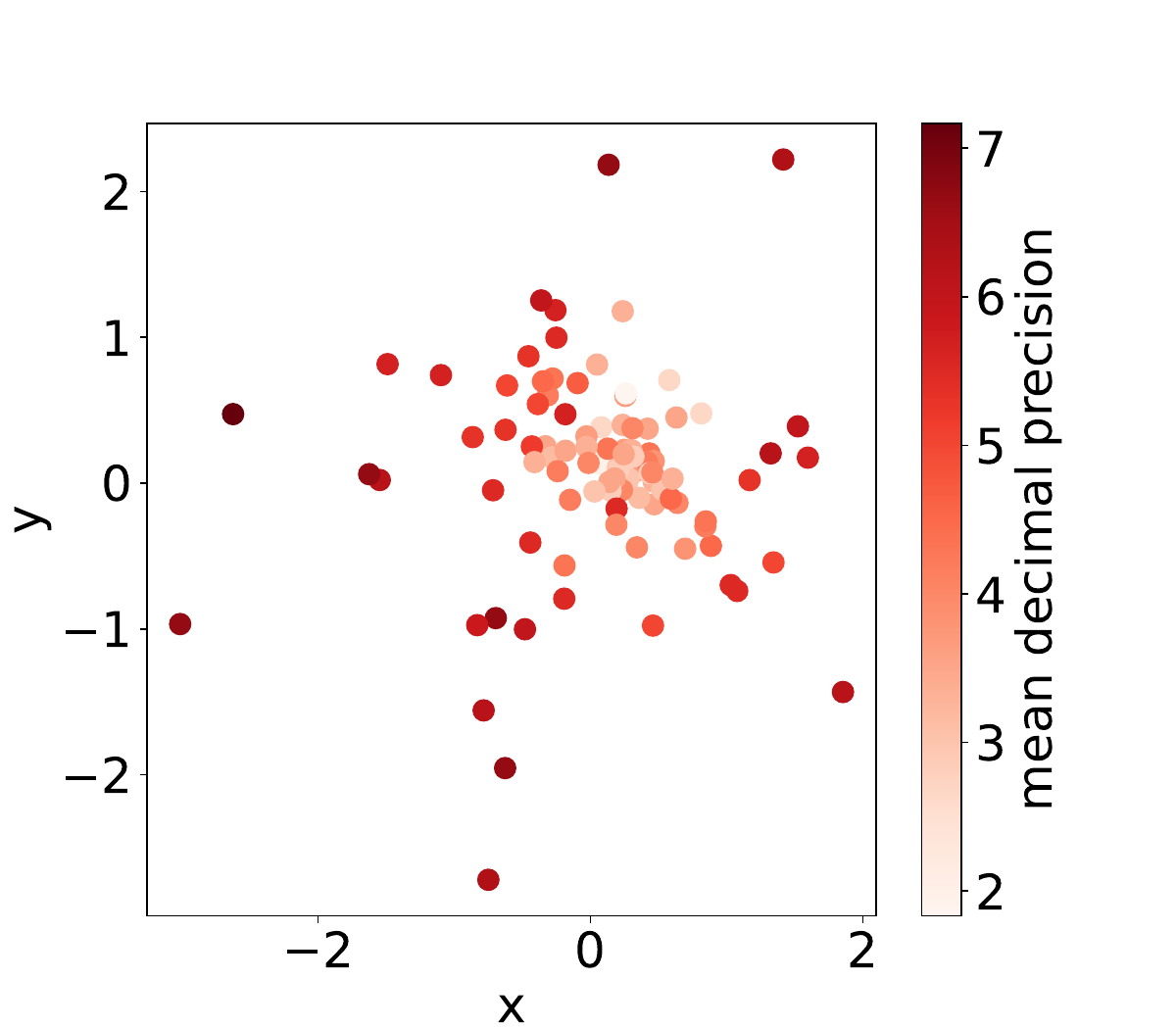}
\includegraphics[width=1\columnwidth]{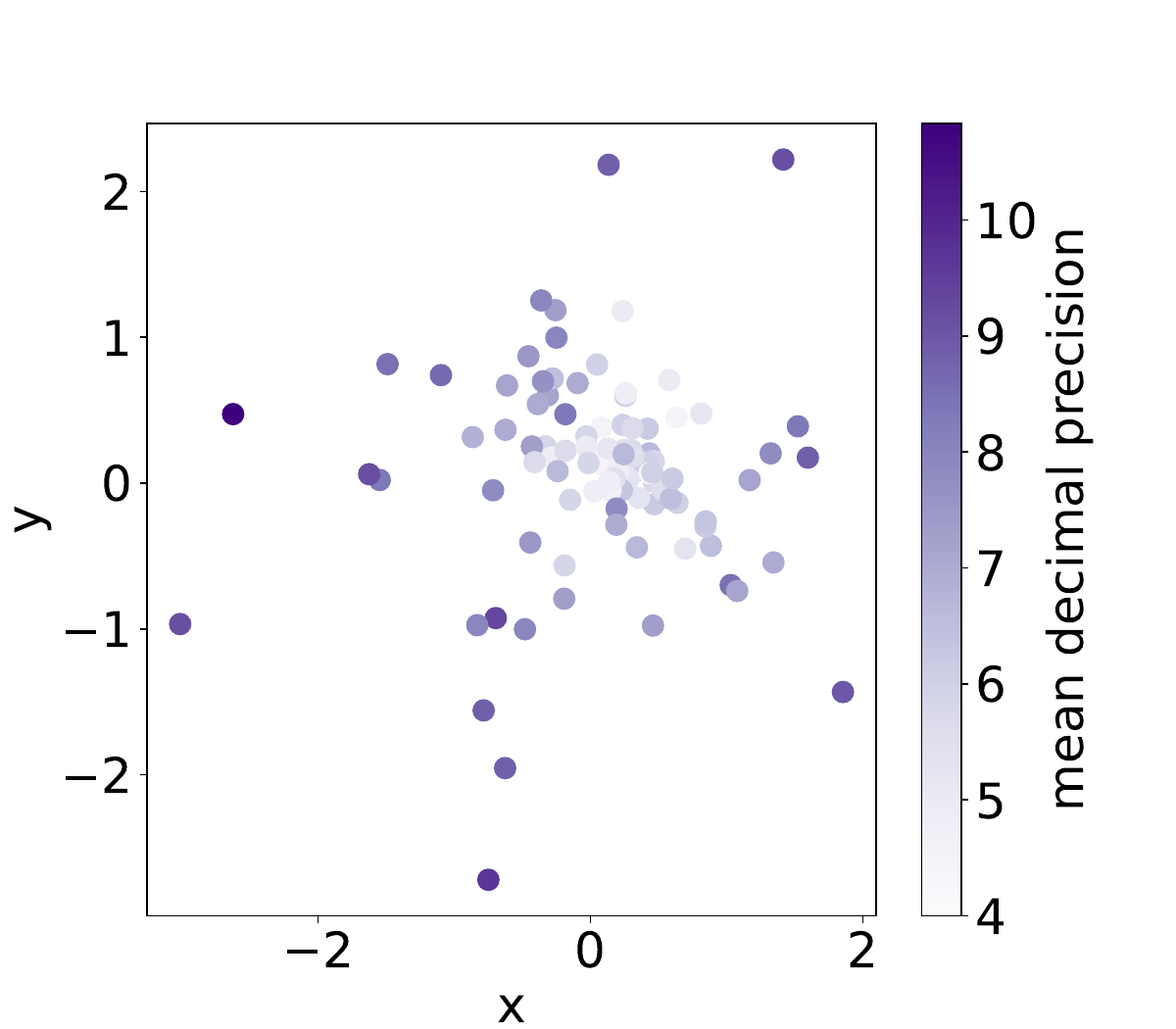}
\caption{Particle positions at $t=6$ for the $N-100$ particle Plummer
  model projected on the x- and y-plane. Colors indicate the mean
  number of accurate decimal places compared to the converged Brutus
  simulation. The left panel gives the result of the calculations
  using Posits (red with 2 to 7 decimal places accurate), while
  Universal to the right (purple, with 4 to 10 decimal places
  accurate). We do not show the fp64 results, as those are almost
  indistinguishable from the results calculated with Universal.
}
\label{fig:N100Plummer_representation}
\end{figure*}

The baseline calculation is performed using Brutus to convergence.  We
subsequently run the same initial realization using fp64.
\Cref{fig:Plummer_N100_precision} shows, as a function of time, the
fraction of model parameters that converge to a certain number of
decimal places (y-axis).  The numerical representation of fp64 drops
linearly with time in the Plummer model over 16 decimal places in 6
N-body time units, losing roughly 3 decimal places per time unit.  For
a final snapshot, this statistics is also presented in
\cref{tab:FinalPrecision} (see also
\cref{fig:N100Plummer_representation}).

Over-plotted is the mean number of decimal places to which the 600
degrees of freedom are represented for the calculations using
Universal. The latter follows the mean of the fp64 simulations,
including the structure in this curve. The most prominent is the
gravothermal collapse of the cluster, around $t=3.4$, leading to a
jump in the phase-space distance curve due to a punctuation event in
the chaotic behavior of the system.

In \cref{fig:N100Plummer_representation}, we present the $N=100$
Plummer model at $t=6$ N-body time units. Both representations look
similar, which is not surprising considering that the number of
accurate decimal places ranges between 2 and 10 (see the color
coding), and \cref{fig:Plummer_N100_precision}.  It is interesting to
notice that the particles in the center are most strongly affected by
precision problems, whereas the particles more to the outskirts tend
to be represented more accurately. This is the case for fp64,
CPPPosits, and Universal (although we only show the latter two).
Universal and fp64 remain more accurate than CPPPosits. The former
managed to keep numerical precision remains between 2 and 7 decimal
places, whereas Universal manages to keeps 4 to 10 decimal places
accurate throughout the calculation; Universal clearly outprforms fp64
for the outliers, but on average both lead to slightly better
performance for fp64 compared to Universal (see
\cref{tab:FinalPrecision}.  We conclude that overall for the average
precision fp64 slightly outperforms Universal, but that Universal is
better in confining the error outliers.

\subsection{Performance comparison}\label{Sect:Performance}

While performing the calculations, we keep track of the efficiency of
the operations in terms of the wall-clock time. In
\cref{tab:EnergyError}, we present an overview of the measured
wall-clock time, and energy conservation of the various calculations.

The 3-body problems are less computationally intensive and suffers
from some overhead, whereas the 100-particle Plummer model is
completely dominated by numerical operations. The difference in
performance between the 3-body and the 100-body problems is probably a
result of the more intensive use of complex operations (in particular
the square root) in the latter (see also
\cite{Murillo2024SquareRootPosit}).

The 100-particle Plummer simulation (see \cref{Sect:N100Plummer}) with
32-bit IEEE-754 are only slightly faster than 64-bit calculations (17
seconds), and 64-bit calculations are about five times faster than the
same 128-bit calculations.  The calculation using 64-bit Universal
took 58\,910 seconds; more than 2000 times slower than fp64. Note that
the Pythagorean and figure-8 problems were comparably slow when
running with the Universal library.

CPPPosit were $\sim 120$ times slower than the same calculations using
floating point arithmetic for the 100-particle Plummer model. The
3-body problems performed better: the 64-bit and 32-bit CPPPosits were
only 25 and 20 times slower than fp64 calculations. Bear in mind,
though, that the calculations with IEEE floating points were
hardware-supported, whereas Posits were performed in compiled C++.

The floating-point unit on the adopted 13th Gen Intel Core i7-1370P
processor used for the presented calculations executes one or two
floating-point operations per clock cycle, compared to about 30 to 100
cycles for a software-implemented operation. A loss of speed of a
factor 20 to 120 then is not so surprising, but it also does not offer
much leeway for the speed advantages of Posits compared to fp64. A
slow-down of a factor $\sim 2\,000$ of the Universal library compared
to fp64 cannot be explained in terms of the hardware support for the
latter.

The performance of a software emulation, however, should not be
interpreted as an indication of the performance that Posits would
achieve in hardware. The Universal library is designed for
flexibility and ease of use: it supports arbitrary word lengths and
exponent sizes through C++ templates and operator overloading, and
represents numbers in multi-limb integer containers (see
\cref{Sect:Posits_vs_Uiveral}). This generality comes at a cost in
speed, and the library is not necessarily optimized for a specific
posit configuration. A similar practice is common in the study of new
arithmetic formats, such as the fp8 formats recently introduced for
machine learning on GPUs, which are often emulated on IEEE-754 hardware
to study their numerical behavior, irrespective of the speed of the
emulation.  Software performance may change considerably with future
optimizations, and a dedicated hardware implementation of the posit
adder, multiplier, divider, and square-root would be considerably
faster than either of the libraries adopted here.  The emphasis of our
study is therefore on the numerical behavior of Posits, and we report
the wall-clock times primarily to inform researchers about the
current state of software support for Posits.

All timings reported in \cref{tab:EnergyError} are for single-core
runs.  The direct-summation force calculation could be parallelized
(for example with OpenMP), but this has little benefit for $N=3$ and
only modest benefit for $N=100$.  Most of our experiments, however,
consist of ensembles of independent runs (different numerical
formats, time-step parameters, and initial realizations), which are
embarrassingly parallel and can be distributed over many cores
without communication, after which the results are collected and
analyzed.  The arbitrary-precision Brutus calculations were run in
parallel (see \cref{Sect:N100Plummer}).

\begin{table*}[h]
\centering
\caption{Performance in terms of speed-up and energy conservation of
  the simulations with various precisions and using a time step
  parameter $\eta=10^{-4}$. The speed-up (second column) is presented
  in terms of the 64-bit float performance ($t_{\rm 64bit}$), which is
  hardware supported. The initial energy for figure-8's problem is
  $E_0 \simeq -1.287$, for the Pythagorean problem $E_0 \simeq
  -12.82$, and for the virialized Plummer model, it is $E_0 \equiv
  -0.25$. The D9 simulations represent an actual astronomical system,
  and has a total energy of $-1.48 \cdot 10^{49}$\,erg.  }
\begin{tabular}{lllllllll}
\hline \hline
& speed-up & figure-8 & Pythagoras & Plummer & D9 \\
& [$t_{\rm fp64}$]& $|dE/E_0|$ & $|dE/E_0|$ & $|dE/E_0|$ & $|dE/E_0|$ \\
\hline
128-bit floats & 0.21 & $ 6.4 \cdot 10 ^{-17}$ & $3.5 \cdot 10 ^{-11}$ & $9.2 \cdot 10 ^{-15}$ & $5.8 \cdot 10^{-8}$ \\
64-bit floats & 1 & $ 7.2 \cdot 10 ^{-13}$ & $8.2 \cdot 10 ^{-8}$ & $2.0 \cdot 10 ^{-10}$ & $5.8 \cdot 10^{-8}$ \\
32-bit floats & 1.05 & $ 1.3 \cdot 10 ^{-4}$ & $5.7 \cdot 10 ^{-1}$ & $1.0 \cdot 10 ^{-5}$ & $5.7 \cdot 10^{-4}$\\
64-bit CPPPosits & 0.008& $ 4.6 \cdot 10 ^{-8}$ & $1.9 \cdot 10 ^{-7}$ & $3.7 \cdot 10 ^{-9}$ & $8.4 \cdot 10^{-9}$ \\
32-bit CPPPosits & 0.010& $ 8.9 \cdot 10 ^{-3}$ & $8.3 \cdot 10 ^{+34}$ & $3.5 \cdot 10 ^{-3}$ & $3.1 \cdot 10^{-1}$ \\
64-bit Universal &0.0004&$7.5 \cdot 10 ^{-13}$ & $1.5 \cdot 10 ^{-7}$ & $5.1 \cdot 10^{-9}$ & $2.9 \cdot 10^{-8}$ \\ 
32-bit Universal &0.0015& $4.3 \cdot 10^{-5}$ & $5.7 \cdot 10 ^{-1}$ &$1.8 \cdot 10^{-4}$   & $8.2 \cdot 10^{-4}$  \\

Brutus converged &$10^{-6}$& $10^{-8}$         & $10^{-8}$            & $10^{-8}$            & $10^{-8}$  \\
\hline \hline
\end{tabular}
\label{tab:EnergyError}
\end{table*}

In \cref{tab:EnergyError}, we present the final error in the total
energy of the N-body systems relative to the initial system's energy
(which in virialized units $E_0 \equiv -0.25$). The energy
conservation for each of the calculations using CPPPosits and
Universal is an order of magnitude worse than for fp64. We do not list
the 16-bit performance because they fail to achieve any scientifically
meaningful result, even when adopting brain floating points.

\section{Conclusions}

We present a series of numerical experiments for validating and
verifying numerical solvers for Newton's chaotic N-body problem.
These tests were applied to IEEE-754 floating point arithmetic and
Posits.  We carried out a comparison with low precision (16-bit,
32-bit) and regular precision (64-bit). All the 16-bit calculations
fail because of their limited dynamic range, and also 32-bit precision
tends to lead to unacceptable errors.  Although sufficient accuracy is
hard to define in this case, we argue that IEEE-754 floating point
arithmetic gives acceptable results for most cases, but more precision
would be preferable.

We conclude that Posits (as implemented in CPPPosits or Universal),
may not be the optimal choice for integrating chaotic or stiff
ordinary differential equations because of concerns about the
occasional large rounding error (CPPPosits), or because of the low
speed of the implementation (Universal).  Our main concerns are:
\begin{itemize}
\item[$\bullet$] All numerical representations have difficulties with
  Galilean invariancy. This aspect is hardy ever tested in actual
  problems. The effect can be rather pronounced when simulating a
  system with a wide dynamic range, or with non-trivial substructures,
  such as a planetary system in a stellar cluster.  The consequences
  of this discrepancy can be quite severe, and Galilean invariancy of
  substructures in large scale calculations are rarely tested, even
  though they may lead to questionable results.
\item[$\bullet$] The implementation of CPPPosits fail to provide
  reliable errors for calculations over a wide dynamic range.  Here
  reliable errors are those that stay withing the expected boundaries;
  CPPPosits occasionally produces errors larger than expected. If
  sufficient precision (in terms of significant bits) are available
  this problem can be effectively addressed. For the same number of
  bits, Universal produces a narrower range of errors compared to IEEE
  floating point arithmetic, but the mean number of preserved correct
  digits is smaller. Too many outliers could be disastrous for
  scientific interpretation more so than a larger average error, and
  we therefore prefer Universal in terms of precision (but certainly
  not in terms of raw speed).
\item[$\bullet$] The performance of Posits depends sensitively on the
  implementation (at least for the two implementations we tried). For
  floating point, this does not pose a problem because of their IEEE
  compliancy. For Posits the lack of a formal standard\footnote{An
  informal standard for Posits can be found at\\
  \url{https://posithub.org/docs/posit_standard-2.pdf}.} seem to lead
  to differences in implementation, and therewith in speed.
\item[$\bullet$] We further argue that CPPPosits fail to resolve the
  intricacies of close encounters in self-gravitating systems.
  Universal performs considerably more consistent.  For most of the
  time, the calculation may well be sufficiently precise for
  scientific interpretation, but once in a while CPPPosits simply fail
  to perform a calculation with sufficient accuracy.  This is most
  notoriously during long integrations of a chaotic system where close
  encounter occur leading to short distances and high velocities.  It
  is in those circumstances the CPPPosits produce relatively large
  errors due to the rounding to the nearest appropriate Posit.  IEEE
  fp64 and Universal behave more predicable.
\item[$\bullet$] The same mathematical representation in Posits may
  round to a different concrete value than in fp64 because the nearest
  representable posit is not the same as the nearest representable
  fp64. The representable numbers of fp64 are uniformly distributed
  in each binade, whereas Posits have a tapered precision that is
  higher than fp64 near unity (58 versus 52 fraction bits for 64-bit
  numbers) and lower for very large or very small numbers (see the
  discussion of \cref{tab:FinalPrecision}).  Rounding to a different number is not
  necessarily worse, but it leads to a different realization of the
  round-off error. For chaotic dynamical systems this has profound
  consequences: the results of fp64 and Posits eventually diverge, and
  only a comparison with a converged solution (here obtained with
  Brutus) can establish which of the two is more accurate.
\item[$\bullet$] The largest errors in direct N-body simulations are
  originate from particles in the central portion of the cluster,
  whereas particles in the outskirts are less affected by numerical
  precision. 
\item[$\bullet$] Although 16-bit arithmetic fails to produce
  scientifically meaningful results, bfloat16 offers improved
  precision. While bfloat16 remains less accurate than fp32 and does
  not yield scientifically valid data, it outperforms standard 16-bit
  arithmetic.
\end{itemize}
Nonetheless, applications may exist where these limitations have no
consequence and where posits do offer a reasonable alternative for
fp64.

Our most critical comment about Posits is their apparent loss of
predictability. For 64-bit IEEE double precision, the error is
predictably around 53 bits or 15th significant digits. But it may
occasionally be better. Posits generally perform better near unity,
but the numerical precision diminishes if one moves away from unity.
During integration, an operation may accidentally be ill resolved by a
posit, making a somewhat larger error than in the other operations.
These relatively larger errors also degrade the solution for the
further calculations. One could consider a hybrid approach where an
independent evaluation monitors the achieved precision, applying a
correction in the case of a problem, or executing all operations in
relative coordinates.  Such approach, however, could be expensive in
terms of computer time.  For 16-bit, 32-bit, and 64-bit floating point
such a hybrid approach does not seem advantageous, but considering
that 128-bit floating point arithmetic is about 5 times slower than
fp64, a hybrid multiple-precision approach may save some time at the
cost of a more complex implementation.

We have to note here that our application, the self-gravitating N-body
systems pose one of the most challenging numerical problems around.
Regular IEEE-754 double precision also fails in many of the raised
issues. Some of our trust in regular double precision may come from
our familiarity and acceptance of its aberrations. Irrespective of the
psychological argument, we currently do not see Posits as an
alternative for floating point arithmetic for self-gravitating N-body
systems.

\section*{Acknowledgments}
It is a pleasure to thank Georgi Gaydadjiev, who, after I presented a valedictory lecture for Henry Bal, informed me of the existence of
Posits. I would also like to thank Lourens Veen for suggesting testing
Galileo invariance, Peter Coveney, for directing me to his work on the
chaotic generalized Bernoulli map, Tjarda Boekholt, for including a figure on the energy error as a function of time step, and Douglas
Heggie for many insightful comments on an earlier version of the
manuscript.

\section*{Software}

The various precision N-body codes used for this project can be found
in \url{https://gitlab.strw.leidenuniv.nl/spz/NBotox}. Brutus is
available on
\url{git@gitlab.strw.leidenuniv.nl:spz/brutus-newton.git}. Initial
conditions are generated using the Astrophysical Multipurpose Software
Environment (AMUSE), which can be found on
\url{https://amusecode.org}. The data produced for this project and
the scripts to generate the figures are available at Zenodo.

\section*{Energy used for the calculations}

The majority of computer resources went into running the arbitrary-precision N-body calculations. Each of the converged N=100 Plummer spheres took about 12 days. We performed a total of 4 such runs,
totaling to about 48 days of computer time. Rounding this number to 50
days to accommodate all the other simulations. These jobs spent a total
of 150\,kWh of energy, which is equivalent to 57\,kg of CO$_2$
emission in the Netherlands, according to
https://calculator.green-algorithms.org.

\end{document}